\PassOptionsToPackage{square,comma,numbers,sort&compress}{natbib}  % https://tex.stackexchange.com/questions/367391/option-clash-for-package-squarenatbib
\documentclass[final,1p,times]{elsarticle}

\usepackage{rotating} % to rotate text 90 deg

\usepackage{natbib} %for \citeauthor and \citeyear to work%

\usepackage{lineno}
\usepackage{smartdiagram}  % flowchart with tikz
\usepackage{enumitem} % a), b), c) items in list
\usepackage{multicol}

\usepackage{etoolbox}  % introduces patch command
\usepackage{siunitx}
\usepackage{framed} % Framing content
 
\usepackage{nomencl} % Nomenclature package
\makenomenclature
\renewcommand*\nompreamble{\begin{multicols}{2}}
\renewcommand*\nompostamble{\end{multicols}}

\renewcommand\nomgroup[1]{% 
  \item[\bfseries
  \ifstrequal{#1}{A}{ Abbrevations }{%
  \ifstrequal{#1}{D}{ Dimensionless Variables}{%
  \ifstrequal{#1}{S}{ Symbols}{%
  \ifstrequal{#1}{B}{ Subscipts}{%
  \ifstrequal{#1}{P}{ Superscipts}}}}}%
]}

\makeindex
\usepackage[T1]{fontenc}
\usepackage[utf8]{inputenc}
\usepackage[english]{babel}
\addto\captionsenglish{} % https://tex.stackexchange.com/questions/17489/change-caption-name-of-figures
\addto\captionsenglish{} %% LLW: If you use babel, there is a mix up with spaces
\usepackage{amsmath, amsthm, amsfonts, amssymb}
\usepackage{mathrsfs} % fancy math letters as is middle ages ;)

\usepackage{bbm}

\usepackage{caption}

\usepackage{graphicx} % dummy figures
\usepackage{subcaption} % two subfigures - one caption

\usepackage{floatrow}
\usepackage{floatpag}
\floatpagestyle{plain}
\rotfloatpagestyle{plain}

\usepackage{setspace}
\usepackage{verbatim}

\usepackage{lscape}
\usepackage{graphicx}
\usepackage{amssymb}
\usepackage{tikz} % TM: 06/11/18

\usepackage{lineno}

\newcommand{\centered}[1]{\begin{tabular}{l} #1 \end{tabular}}  % GG - table: ADE squeares
\usepackage{multirow} % GG  - table: ADE squeares

\usepackage{booktabs}

\usepackage{hyperref} % Enable clickable links
\hypersetup{urlcolor=blue, 
    citecolor=blue,
    filecolor=blue,
    linkcolor=blue,
    urlcolor=blue,
colorlinks=true} % Colors hyperlinks in blue - change to black if annoying

\usepackage{acro}
\DeclareAcronym{LB}{
    short = LB,
    long  = lattice Boltzmann,
    class = nomencl
    }
    
\DeclareAcronym{LBM}{
    short = LBM,
    long  = lattice Boltzmann method,
    class = nomencl
    }
    
\DeclareAcronym{FEM}{
    short = FEM,
    long  = finite element method,
    class = nomencl
    }    
    
\DeclareAcronym{NS}{
    short = NS,
    long  = Navier Stokes,
    class = nomencl
    }
    
\DeclareAcronym{ADE}{
    short = ADE,
    long  = advection-diffusion equation,
    class = nomencl
    } 
        
\DeclareAcronym{CM}{
    short = CM,
    long  = central moments,
    class = nomencl
    }    

\DeclareAcronym{SRT}{
    short = SRT,
    long  = single relaxation time,
    class = nomencl
    } 

\DeclareAcronym{MRT}{
    short = MRT,
    long  = multiple relaxation time,
    class = nomencl
    } 
    
\DeclareAcronym{TRT}{
    short = TRT,
    long  = two relaxation time,
    class = nomencl
    } 

\DeclareAcronym{BC}{
    short = BC,
    long  = boundary condition,
    class = nomencl
    }                 

\DeclareAcronym{BB}{
    short = BB,
    long  = bounce-back,
    class = nomencl
    }  
    
\DeclareAcronym{EQ}{
    short = EQ,
    long  = equilibrium scheme,
    class = nomencl
    }  
    
\DeclareAcronym{ABB}{
    short = ABB,
    long  = anti-bounce-back,
    class = nomencl
    }  
            
\DeclareAcronym{IBB}{
    short = IBB,
    long  = interpolated-bounce-back,
    class = nomencl
    }
    
\DeclareAcronym{IABB}{
    short = IABB,
    long  = interpolated-anti-bounce-back,
    class = nomencl
    }

\DeclareAcronym{Re}{
    short = $Re$,
    long  = Reynolds number,
    class = nomencl
    }
\DeclareAcronym{Pr}{
    short = $Pr$,
    long  = Prandtl number,
    class = nomencl
    }
\DeclareAcronym{Nu}{
    short = $Nu$,
    long  = Nusselt number,
    class = nomencl
    }               

\nomenclature[A]{$LB$}{\acl{LB}}
\nomenclature[A]{$LBM$}{\acl{LBM}}
\nomenclature[A]{$FEM$}{\acl{FEM}}
\nomenclature[A]{$SRT$}{\acl{SRT}}
\nomenclature[A]{$MRT$}{\acl{MRT}}
\nomenclature[A]{$TRT$}{\acl{TRT}}
\nomenclature[A]{$BC$}{\acl{BC}}
\nomenclature[A]{$CM$}{\acl{CM}}
\nomenclature[A]{$ADE$}{\acl{ADE}}
\nomenclature[A]{$NS$}{\acl{NS}}
\nomenclature[A]{$IBB$}{\acl{IBB}}
\nomenclature[A]{$IABB$}{\acl{IABB}}
\nomenclature[A]{$BB$}{\acl{BB}}
\nomenclature[A]{$EQ$}{\acl{EQ}}

\nomenclature[S]{$f_i$}{i-th distribution function for velocity field}
\nomenclature[S]{$h_i$}{i-th distribution function for internal energy field}

\nomenclature[S]{$g$}{distribution function of interest}

\nomenclature[S]{$\nu$}{kinematic viscosity}
\nomenclature[S]{$\mu$}{dynamic viscosity}

\nomenclature[S]{$lu$}{lattice unit}

\nomenclature[S]{$L$}{length of the domain}
\nomenclature[S]{$D$}{diameter}
\nomenclature[S]{$H$}{internal energy}
\nomenclature[S]{$\mathbb{M}$}{transformation matrix}
\nomenclature[S]{$\mathbb{N}$}{shift matrix}
\nomenclature[S]{$\mathbb{S}$}{relaxation matrix}

\nomenclature[S]{$\mathbf{x}$}{position vector}
\nomenclature[S]{$\mathbf{e}$}{characteristic lattice velocity}
\nomenclature[S]{$q$}{number of discrete velocities}
\nomenclature[S]{$\mathbf{u}$}{macroscopic velocity vector}
\nomenclature[S]{$\mathbf{\xi}$}{continuous particle velocity space}
\nomenclature[S]{$\rho$}{fluid density}
\nomenclature[S]{$\mathscr{C}$}{forward cumulant transform}
\nomenclature[S]{$c$}{cumulant of distribution function}
\nomenclature[S]{$\Upsilon$}{raw moment of distribution function}
\nomenclature[S]{$\tilde{\Upsilon}$}{central moment of distribution function}
\nomenclature[S]{$k$}{thermal conductivity}
\nomenclature[S]{$c_s$}{lattice speed of sound}
\nomenclature[S]{$N$}{number of spatial dimensions}
\nomenclature[S]{$s_i$}{i-th relaxation frequency}

\nomenclature[S]{$\Omega$}{collision operator}
\nomenclature[S]{$\mathscr{W}$}{the transformation to {\it moments}, {\it central moments} or {\it cumulants}}
\nomenclature[P]{$eq$}{equilibrium}
\nomenclature[P]{$\backsim$}{quantity in central moment space}
\nomenclature[P]{$\star$}{post collision quantity}
\nomenclature[P]{$H$}{quantity related to internal energy field}

\nomenclature[D]{$Re$}{\acl{Re}}
\nomenclature[D]{$Pr$}{\acl{Pr}}
\nomenclature[D]{$Nu$}{\acl{Nu}}
\newcommand{\pr}[1]{\frac{\partial}{\partial #1}}

\newcommand{\eq}{\text{eq}}

\def\etal.{et\penalty50\ al.}

\usepackage{etoolbox}
\newcommand{\QF}{\ensuremath{\text{Q}_\text{F}}}
\newcommand{\QH}{\ensuremath{\text{Q}_\text{H}}}
\newcommand{\DQQ}[3]{%
	\ifstrempty{#3}{%
			\ensuremath{\text{D}#1\QF #2}%
	}{%
		\ifstrempty{#2}{%
			\ensuremath{\text{D}#1\QH #3}%
		}{%
			\ensuremath{\text{D}#1\QF #2\QH #3}%
		}%
	}%
}
\newcommand{\DQ}[2]{\ensuremath{\text{D}#1\text{Q}#2}}

\usepackage[capitalise]{cleveref}
\journal{XXX} %

\begin{document}
\begin{frontmatter}


\title{A comparative study of 3D Cumulant and Central Moments lattice Boltzmann schemes with interpolated boundary conditions
for the simulation of thermal flows in high Prandtl number regime
}



%\author{G. Gruszczyński}

%\address{California, United States}

\author[First,Second]{G. Gruszczy\'nski \corref{cor1}%\fnref{label2}
}
\ead{ggruszczynski@gmail.com}

\author[Third,First]{\L{}. \L{}aniewski-Wo\l{}\l{}k}
%\ead{autor2@cea-ifac.es}
%\ead[url]{www2.cea-ifac.es}

%\fntext[label2]{Nota al pie para el autor 1}
\cortext[cor1]{Corresponding author.}

\address[First]{Institute of Aeronautics and Applied Mechanics, Warsaw University of Technology, Warszawa, Poland}
\address[Second]{Interdisciplinary Centre for Mathematical and Computational Modelling, University of Warsaw, Warszawa, Poland}
\address[Third]{School of Mechanical and Mining Engineering, The University of Queensland, St Lucia, Australia}

\begin{abstract}
Thermal flows characterized by high Prandtl number are numerically challenging as the transfer of momentum and heat occurs at different time scales.
To account for very low thermal conductivity and obey the Courant-Friedrichs-Lewy condition, the numerical diffusion of the scheme has to be reduced.
As a consequence, the numerical artefacts are dominated by the dispersion errors commonly known as \textit{wiggles}.
In this study, we explore possible remedies for these issues in the framework of \acl{LBM} by means of applying novel collision kernels, lattices with large number of discrete velocities, namely \DQ{3}{27}, and a second-order boundary conditions.

For the first time, the cumulant-based collision operator is utilised to simulate both the hydrodynamic and the thermal field.
Alternatively, the advected field is computed using the central moments' collision operator.
Different relaxation strategies have been examined to account for additional degrees of freedom introduced by a higher order lattice.

To validate the proposed kernels for a pure advection-diffusion problem, the numerical simulations are compared against analytical solution of a Gaussian hill.
The structure of the numerical dispersion is shown by simulating advection and diffusion of a square indicator function.
Next, the influence of the interpolated boundary conditions on the quality of the results is measured in the case of the heat conduction between two concentric cylinders.
Finally, a study of steady forced heat convection from a confined cylinder is performed and compared against a Finite Element Method solution.

It is known from the literature, that the higher order moments contribute to the solution of the macroscopic advection-diffusion equation.
Numerical results confirm that to profit from lattice with a larger number of discrete velocities, like \DQ{3}{27}, 
it is not sufficient to relax only the first-order central moments/cumulants of the advected field.
In all of the performed benchmarks, the kernel based on the two relaxation time approach has been shown to be superior or at least as good as counter-candidating kernels.
\end{abstract}


\end{frontmatter}

\noindent\fbox
{
    \parbox{\textwidth}
    {\printnomenclature}
}

%\linenumbers


\acresetall % reset acronyms
\section{Introduction}\label{sec:Intro}
The \ac{ADE} may be viewed as a key building block in CFD modelling.
In its simplest form, it describes the transport of scalar fields such as temperature, or concentrations of different chemical components across the spatial domain.
The equation can be extended by addition of a source terms to account for chemical reactions, or heat generation.
Once the density of a medium is introduced, the energy transfer can be properly modelled.
Finally, a much more complex, nonlinear \ac{NS} equation describe the motion of a fluid.

Several numerical methods are currently used across both academia and industry for solving both the \acl{NS} and advection-diffusion equations.
These include, but are not confined to, finite difference, finite volume~\cite{versteeg2007introduction}, and \ac{FEM}~\cite{zienkiewicz2005finite}, which all discretise the macroscopic equations directly.
Alternatively to solving discretised macroscopic equations,
the \ac{LBM} focuses on the evolution of a distribution function.
The distribution function define the density of some macroscopic physical quantity (such as mass or chemical concentration) across space. 
From the microscopic perspective, it can be viewed as the probability of finding a quantity-carrying particle with a given velocity and coordinates at a given time.
In the discretised form, the set of velocities is finite.
In each iteration, the discretised distributions are passed to neighbouring nodes along these velocities, in a step called streaming. 
This requires exchange of information with neighbouring nodes.
Next, a local collision step is performed in each node, which relaxes the distribution function toward some form of an equilibrium.

During the last decades, a variety of collision operators have been proposed.
For the detailed discussion of this subject, the reader is referred to the comparative works~\cite{Luo2011,Coreixas2019,Coreixas2020} and established textbooks~\cite{Kruger2017,SauroSucci2019}.
In the recent years, a substantial effort was made to use the mutual independence of the observable quantities (such as density, momentum, stress tensor components, etc.) to separate their relaxation frequencies and improve stability.
In the simplest form of the collision operator, a \acf{SRT} is assigned to all distribution functions.
The next concept follows from the fact that the observable quantities can be found as moments of the distribution function~\cite{d1992generalized,Kruger2017}.
As different relaxation rates can be assigned to each moment, the \ac{MRT} name has been coined~\cite{d1992generalized}.
The moments can be calculated in stationary or moving reference frame and are refereed as raw or \ac{CM} respectively~\cite{Geier2006}.
A method, in which the relaxation rates of odd and even raw moments are separated has been originally proposed by~\citet{Ginzburg2005}, and is known as \ac{TRT}.

The proper choice of the combination of moments used during the collision step plays important role for the stability of the scheme.
~\citet{Dubois2015} have benchmarked two frequently used sets in the contexts of \ac{NS} equations.
Some researchers argue that the orthogonalisation of the transformation from distribution function to the moments' space shall limit the {\it cross-talk} between the moments~\cite{d2002multiple}.
On the other hand~\citet[Appendix I]{Geier2015a} stated that the orthogonalisation process can be the source of spurious couplings.
The article proposed that a more complex, non-linear, statistically independent quantities known as cumulants can be calculated from the distribution function.
Their use in the \ac{LBM} framework resulted in the formulation of the Cumulant Lattice Boltzmann Method, which exhibited very promising numerical properties~\cite{Geier2015a,Geier2017a}.

The \acf{Pr} defines the ratio of kinematic viscosity to thermal diffusivity. 
Modelling high \ac{Pr} flows rises numerical difficulties due to different time scales of the physical phenomena occurring in the hydrodynamic and thermal field. 
Briefly, three distinct approaches to model thermal flows within the \ac{LBM} framework can be distinguished in the literature~\cite{LiLuo2016}.

In a \textit{fully coupled double distribution function} approach,
a second distribution function evolving between the same lattice nodes, at the same timesteps as the hydrodynamic one,
is introduced to simulate the evolution of the energy field~\cite{Doolen1998}.
Due to tight coupling, many benchmarks published for this type of \ac{LBM} models are restricted to \ac{Pr} $ \leq 100$ ~\cite{Mabrouk2020,Du2021,Chen2021} 
or \ac{Pr}$\simeq 1$~\cite{Doolen1998,Guo2007,Feng2015,Huang2015a,Li2016,Chen2017,Sharma2017,Sharma2018,Fei2018c,Fei2018d,Lu2018,Hosseini2018,Feng2018,Xu2019,Yip2019,Sajjadi2019}.
Limits of this way of modelling, are explored in the current paper.

Another approach, called multispeed \ac{LBM} uses single distribution function, but utilises a lattices with a large velocity set~\cite{Nie2008}.
The expanded velocity set allows to eliminate the aliasing of the higher order moments of the distribution.
As a result, the internal energy can be incorporated as a quantity solved by \ac{LBM}.
This is not possible for standard lattices, as the limited number of independent moments results in a fixed relation between the internal energy and the pressure, which is referred as \textit{isothermal \ac{LBM}}.
The use of the multispeed method with the SRT collision operator would lead to the \ac{Pr} being a constant of the model, 
as the relaxation rate corresponding to the thermal conductivity could not be set independently of from the kinematic viscosity~\cite{He1998a,Cercignani1988}.
Successful decoupling has been achieved by~\citet{Chen2014a} followed by~\citet{Shan2019}. 
A collision operator with multiple relaxation frequencies in a moving reference frame has been proposed and validated for \ac{Pr} $\in\{ 0.5, 1, 2 \}$.

In the hybrid approach, two distinct numerical methods are used to solve the hydrodynamics and the advected field.
Examples of this approach include studies, where \ac{LBM} solver was used for the \ac{NS} equations and a finite difference~\cite{Lallemand2003,Saito2021} or finite volume~\cite{Feng2018} method for the thermal field.

Finally, some researches attempt to take advantage of high \ac{Pr} regime to decouple \ac{NS} and \ac{ADE} solvers.
In this method, both fields can be computed using \ac{LBM} solver, but on separated lattices~\cite{Parmigiani2009,Nabavizadeh2021}.
\citet{Parmigiani2009} notes that either the timestep or grid spacing of the second distribution can be decoupled.
Although the spatial decoupling demands interpolations between lattices, the overall computational cost is reduced.
The group managed to simulate flow with \ac{Pr} up to $1000$.
However, when the heat transport becomes velocity controlled, the advantages of decoupling deteriorate~\cite{Parmigiani2009}.

Discretisation of the velocity space in the Lattice Boltzmann Method corresponds to the number of links between each of the neighbouring nodes in the lattice.
In general, the choice of the discrete velocity set is dictated by the target macroscopic equation.
On the one hand, large velocity sets are expected to provide more accurate~\cite{Feng2015} and stable solutions~\cite{Suga2006}.
On the other hand, they contribute to increased memory consumption, computational and communication cost, as well as complexity of boundary conditions.
The velocity sets are commonly described using \DQ{d}{q} notation,
where $d$ corresponds to the number of spatial dimensions and $q$ accounts for the number of discrete velocities.
The \QF{} and \QH{} subscripts will be used in the present work to distinguish the lattices used for hydrodynamic and thermal distribution functions, hence defining the lattice as \DQQ{d}{x}{y}.

To solve the advection-diffusion equations, a lower order lattices like \DQQ{2}{}{5}~\cite{Sharma2017,Sharma2018,Fei2018c,Lu2018} 
and \DQQ{3}{}{7}~\cite{Yoshida2010,Li2014,Yoshida2014,Karani2015,Chen2016,Chen2017b,Xu2019,Sajjadi2019} are frequently used.
However, to properly recover complex thermal flows (e.g. measure the critical Rayleigh number), a higher-order lattice is required~\cite{Feng2015}.
According to~\citet{Huang2011}, the accuracy of \DQQ{2}{}{4}, \DQQ{2}{}{5} and \DQQ{2}{}{9} lattices can be comparable, but the effects of low diffusivity were not considered in their study.
Additionally, lattices such as \DQ{3}{15} shall be avoided, as it is not feasible to derive the shift matrix from the raw moment space (discussed in next section) following the procedure presented by~\citet{Asinari2008}.
Apart from increasing the number of discrete velocities, 
it is possible to take advantage of a specific geometry of the problem by either scaling a lattice into cuboidal one~\cite{Yahia2021} 
or by adding a correction terms to account for axisymmetric flows~\cite{Hajabdollahi2019}.
Interestingly, when the relaxation frequency is exactly equal to one, a memory efficient LBM scheme can be implemented, by eliminating the storage of the distribution function.
From the theoretical perspective, a memory savings of up to $\sim 86 \%$ can be achieved for the D3Q27 lattice~\cite{Matyka2021}.

In the current study, a linear interpolation scheme~\cite{HamedBouzidi2001} is implemented to represent a boundary located between lattice nodes and improve the representation of the geometry.
Readers interested in a more detailed studies on the application of an interpolated-anti-bounce-back \acl{BC} are refereed to~\cite{Dubois2009,Ginzburg2009viscosity,Li2013b,Dubois2019}.

Application of standard lattices, such as \DQQ{2}{}{9}~\cite{Elseid2018} and \DQQ{3}{}{15}~\cite{Hajabdollahi2018}, -19 or -27, raises the question of the appropriate relaxation of higher-order moments.
Due to its simplicity, the SRT collision operator is frequently applied~\citep{Guo2007,Li2012,Huang2013,Huang2015a,Grucelski2015,Chen2016,Feng2015,Karani2015,McCullough2016,Lu2017,McCullough2018a,Mabrouk2020,Du2021}.
When it comes to use a more advanced \ac{MRT}~\citep{Yoshida2010,Wang2013,Yoshida2014,Huang2015a,Li2016,Chai2016,Cui2016,Sajjadi2019,Xu2019} or \ac{CM}~\citep{Sharma2017,Sharma2018,Elseid2018,Hajabdollahi2018,Fei2018c,Fei2018a,Hajabdollahi2019} collision operator, various relaxation strategies can be identified.
Some authors~\cite{Yoshida2010,Yoshida2014,Li2016,Chai2016,Hajabdollahi2018,Elseid2018,Sajjadi2019} relax first-order moments only, setting the rest of moments to their corresponding equilibrium values.
The \ac{TRT} approach was adopted to moment space~\cite{Wang2013,Huang2015a,Xu2019} and central moment space in~\cite{Sharma2018,Fei2018c,Fei2018a}, however in~\cite{Fei2018a,Hajabdollahi2018,Elseid2018} the relaxation of higher-order moments is said to be tunable.

It has been shown by different authors~\cite{Nie2008,Fei2018,Fei2018a,DeRosisLuo2019,gruszczynski2019cascaded}, that truncation of the equilibrium distribution function deteriorates Galilean invariance of the flow model.
\citet{Chopard2009} have discussed the role of the second-order velocity terms in the equilibrium distribution function on the error of the recovered macroscopic equation and proposed a correction term.
\citet{Nie2008} noticed that the error appears as spurious dependence of the macroscopic diffusion coefficient (i.e. viscosity or thermal conductivity) on velocity.
A similar effect has been observed by~\citet{Fei2018a} who used a cascaded collision operator with full order equilibrium to show that thermal diffusivity is independent of Mach number as opposed to works where \ac{MRT} with truncated equilibrium~\cite{Liu2016,Cui2016} were used.
To avoid the above-mentioned issues, a full order equilibrium distribution function is used in the present work.

The~\cref{tab:thermal_lit_digest} presents a concise summary of the models present in the literature, to put the collision operators chosen in this work in a context.
Features such as the type of collision kernel, order of equilibrium, utilized lattice and the range of reported \acl{Pr}s has been extracted.
The publications were usually focused only on selected properties from the aforementioned set.
Usage of SRT collision and truncated equilibrium distribution is prevalent, as is the relaxation of only first raw, or central moments.
It is expected that to some extend, the models listed in~\cref{tab:thermal_lit_digest} would work for \acl{Pr}s other than reported.
\begin{table}[htbp]\centering
\begin{tabular}{cl@{}clclr}
\toprule
Year & Reference & Notes & Collision & $h^\text{eq}$ ord. & Lattice & \acs{Pr} \\ \midrule
\citeyear{Doolen1998}	& \citet{Doolen1998}     	& & SRT & $2^{nd}$ & D2Q9 & 0.25$\div$0.71\\
\citeyear{Guo2007} 		& \citet{Guo2007} 		    && SRT & $2^{nd}$ & D2Q9 & 0.71$\div$1 \\
\citeyear{Parmigiani2009} & \citet{Parmigiani2009} 	&(a)& SRT & $1^{st}$ & D2Q5 & 10$\div$1000 \\
\citeyear{Yoshida2010} 	& \citet{Yoshida2010} 		&& MRT-$1^{st}$ & $1^{st}$ & D3Q7  & --- \\
\citeyear{Huang2013}	& \citet{Huang2013} 		&& SRT & $2^{nd}$ & D2Q9 & 0.02 \\
\citeyear{Karlin2013}	& \citet{Karlin2013}		&& Entropic & $2^{nd}$ & DdQq & 1 \\
\citeyear{Wang2013}   & \citet{Wang2013}     		&& MRT-TRT & $1^{st}$ & D2Q5 & 0.71$\div$7 \\
\citeyear{Chen2014a}	& \citet{Chen2014a}			&(b)& CM-TRT & $4^{th}$ & D2Q27 & 0.5$\div$2 \\
\citeyear{Yoshida2014}	& \citet{Yoshida2014}  	 	&& MRT-$1^{st}$ & $1^{st}$ & D3Q7  & --- \\
\citeyear{Huang2015a}	& \citet{Huang2015a}   		&& MRT-TRT & $2^{nd}$ & D2Q9   & 0.2 \\   
\citeyear{Feng2015}		& \citet{Feng2015} 			&& SRT & $2^{nd}$ & D3Q15/19/27 & 0.71 \\	
\citeyear{Li2016}		& \citet{Li2016} 		    &(c)& MRT-$1^{st}$& $1^{st}$ & D3Q7 & 0.71 \\
\citeyear{Sharma2017}   & \citet{Sharma2017}     	&& CM-TRT & $\infty$ & D2Q5 & 0.71 \\
\citeyear{Xu2017}   & \citet{Xu2017}     			&(e)& MRT-TRT & $1^{st}$ & D2Q5 & 0.71 \\
\citeyear{Lu2017}	& \citet{Lu2017}                && SRT & $2^{nd}$  & D2Q9 & 1 \\
\citeyear{Sharma2018}   & \citet{Sharma2018}     	&& CM-TRT & $\infty$ & D2Q5 & 0.71 \\
\citeyear{Fei2018d}		& \citet{Fei2018d} 			&& SRT & $2^{nd}$ & D2Q9  & 0.71 \\
\citeyear{Feng2018}		& \citet{Feng2018} 			&& SRT & $2^{nd}$ & D2Q9  & 0.71 \\
\citeyear{Hajabdollahi2018} & \multicolumn{2}{l}{\citet{Hajabdollahi2018}} & CM-$1^{st}$ & $\infty$ & D3Q15  & 0.71 \\
\citeyear{Elseid2018} & \citet{Elseid2018}       	&& CM-$1^{st}$ & $\infty$ & D2Q9 & 0.71$\div$1 \\
\citeyear{Fei2018c} & \citet{Fei2018c}         		&(d)& CM-TRT & $\infty$ & D2Q5  & 0.71 \\
\citeyear{Lu2018}	& \citet{Lu2018}                && MRT-TRT & $2^{nd}$  & D2Q9 & 0.7$\div$1\\
\citeyear{Hosseini2018} & \multicolumn{2}{l}{\citet{Hosseini2018}} & SRT & $1^{st}$ &  D2Q9 & 0.7$\div$1 \\
\citeyear{Shan2019} & \citet{Shan2019}         		&(b)& CM-TRT & $9^{th}$ & D2Q37  & 0.5\\
\citeyear{Xu2019} & \citet{Xu2019}             		&& MRT-TRT  & $1^{st}$ & D3Q7  & 0.7$\div$7\\
\citeyear{Sajjadi2019} & \citet{Sajjadi2019}   		&(c)& MRT-$1^{st}$ & $1^{st}$ & D3Q7  & 0.73\\
\citeyear{Lu2019}	& \citet{Lu2019}                && MRT-TRT & $2^{nd}$  & D2Q9/D3Q7 & 0.02 \\
\citeyear{Yip2019} & \citet{Yip2019}                && SRT & $1^{st}$ & D2Q5 & 0.2 \\
\citeyear{Mabrouk2020} & \citet{Mabrouk2020}    	&& SRT & $1^{st}$ & D2Q9 & 50 \\
\citeyear{Du2021} &  \citet{Du2021}                 && SRT & $2^{nd}$ &  D2Q9 & 50 \\
\citeyear{Chen2021} & \citet{Chen2021}         		&& SRT  & $1^{st}$ & D3Q7 & 100 \\
\citeyear{Nabavizadeh2021} & \citet{Nabavizadeh2021}&(a)& SRT  & $1^{st}$ & D2Q9 & 0.01$\div$100 \\
2022 				& current model     		 	&& CM-TRT & $\infty$  & D3Q27 & 10$\div$1000 \\ 
\bottomrule
\end{tabular}
%} % resize box
\caption{The summary of models present in the literature. 
$h^\text{eq}$ ord. - Order of equilibrium distribution function used for the advection-diffusion equation (value $\infty$ means that equilibrium is not truncated, 
\textbf{Pr} - \acl{Pr}. 
Notes: a - NS and ADE were computed on decoupled lattices, 
b - multispeed model, 
c - collision based on~\cite{Yoshida2010}, 
d - collision based on~\cite{Cui2016},
e - collision based on~\cite{Wang2013}.
The $\div$ sign is used to indicate the range of \acs{Pr} for which the model has been benchmarked.
The --- sign means that the definition of \acs{Pr} does not apply, i.e. the model was presented for advected field only.
}
\label{tab:thermal_lit_digest}
\end{table}

In this contribution, the role of different relaxation approaches for the advection-diffusion problems is investigated.
To the best of the authors' knowledge a cumulant collision kernel is applied to simulate the advection-diffusion equation for the first time.
A set of benchmarks with different complexity has been conducted, to isolate the factors which can affect the numerical simulations.
For the set of numerically investigated cases, specified for a range of non-dimensional numbers, the relaxation of higher-order moments has to be adjusted to achieve the benefits of lattice with a large stencil. 
Concluding, current work is focused on the numerical limits of the central moment and cumulant collision operators applied to the variant of advection-diffusion equation, namely transport of internal energy in a homogeneous, isotropic medium with first and second-order boundary conditions using a fully coupled double distribution function approach on a \DQQ{3}{27}{27} lattice.

This paper is organised as follows. 
In the \cref{sec:Model_description}, the \ac{LBM} routine and the investigated collision kernels are described.
In \cref{sec:BC}, the interpolated boundary conditions, which allow to better represent walls location, are presented.
The validation and tests of the model is discussed in \cref{sec:benchmarks}.
The future outlook is given in \cref{sec:future_outlook}.
Finally, the summary and conclusions are formulated in \cref{sec:conclusions}.
For clarity of the manuscript, some of the derivations regarding discretisation of the distribution function, two relaxation time approach, the transformation matrices and the source treatment have been listed in~\ref{app:edf_discretization},~\ref{app:TRT},~\ref{app:NM_matrices} and~\ref{app:source_term} respectively.

\acreset{BB,BC,IBB,IABB,ABB,EQ,Re,Pr,Nu}
\section{Model description}\label{sec:Model_description}
In the present study, the flow of fluid and internal energy balance is modelled by the Navier-Stokes and advection-diffusion equation respectively.
Only one-way coupling between these equations will be considered.
The hydrodynamics is described by the continuity and momentum equations,
\begin{linenomath}\begin{align}
\frac{\partial \rho}{\partial t} + \nabla \cdot \rho \textbf{u} &= 0, \label{eq:continuity} \\
\rho\left( \frac{\partial \textbf{u} }{\partial t} + (\textbf{u} \cdot \nabla) \textbf{u}\right)
&=
-\nabla p + \nabla \cdot (\mu[ \nabla \textbf{u} + (\nabla \textbf{u})^\top]), \label{eq:NS} % + \textbf{F}_b, 
\end{align}\end{linenomath}
where $\textbf{u}$ is the fluid velocity, $\rho$ is the density, $p$ is pressure, and $\mu$ is the viscosity coefficient.
Although the \ac{LBM} is a weakly compressible method, it is used in this study in the incompressible regime ($\rho\simeq\text{const}$).

The heat transfer is described using an internal energy field expressed as $H = \rho c_v T$, where $c_v$ is a specific heat capacity at constant volume and $T$ is the temperature.
Omitting viscous heat production and assuming that the flow is incompressible, the conservation of internal energy can be written as~\cite{Shi2004,LiLuo2016,zikanov2019essential},
\begin{linenomath}\begin{align}
\dfrac{\partial}{\partial t} (\rho c_v T ) + \nabla \cdot (\boldsymbol{u} \rho c_v T ) &= \nabla \cdot (k \nabla T), %\label{eq:internal energy_transfer_eq}
\end{align}\end{linenomath}
with thermal conductivity of the fluid being denoted by $k$.
As the conjugate heat transfer in the surrounding medium is not considered in this study, the $c_v$ is set to $1$.
Having calculated the velocity field from \cref{eq:continuity,eq:NS}, the advection-diffusion of $H$ can be solved independently.
Readers interested in the Chapman-Enskog procedure for the advected field are referred to the work of~\citet{Shi2004}.

\subsection{Lattice Boltzmann Method}
The basic principles of the~\ac{LBM} can be found in a book by~\citet{Kruger2017} or~\citet{SauroSucci2019}.
This section will present the relevant details of the procedure, needed to understand the differences between different collision operators used.

The fluid flow and temperature field in the present~\ac{LBM} framework is described with two distributions functions, $f_\alpha$ and $h_\alpha$, with the corresponding set of velocity vectors $\mathbf{e}_\alpha$.
The connection between $\mathbf{f}$ and $\mathbf{h}$, and macroscopic flow fields is described by,
\begin{linenomath}\begin{align} 
\rho &= \sum_\alpha f_\alpha, \label{eq:f_0th_mom}\\
\rho \mathbf{u} &=  \sum_\alpha \mathbf{e}_\alpha f_\alpha,  \label{eq:f_1st_mom}\\
\rho c_v T &= \sum_\alpha h_\alpha.	\label{eq:h_0th_mom}
\end{align}\end{linenomath}
The general evolution equation of these distributions can be decomposed into two steps: collision and streaming.
The collision is a nonlinear operator $\Omega$ acting on the distribution function at a specific time and location,
while the streaming spreads this distribution function along the velocity vectors $\mathbf{e}_\alpha$.
This is expressed as,
\begin{linenomath}\begin{align}
f_\alpha(\mathbf{x} + \mathbf{e}_\alpha\delta t, t + \delta t ) &= \Omega_{\text{F},\alpha} \left(\mathbf{f}(\mathbf{x}, t) \right),\\
h_\alpha(\mathbf{x} + \mathbf{e}_\alpha\delta t, t + \delta t ) &= \Omega_{\text{H},\alpha} \left(\mathbf{h}(\mathbf{x}, t) \right).
\end{align}\end{linenomath}

The formulations of the \ac{LBM} are presented for the \DQ{3}{27} lattice, as restrictions to smaller lattice or to two dimensions can be easily derived using the presented formulas.
Using the Euclidean basis, the discrete velocities for a \DQ{3}{27} lattice read,
\begin{linenomath}\begin{align}
\textbf{e} &= [\textbf{e}_{\cdot x}, \textbf{e}_{\cdot y}, \textbf{e}_{\cdot z}], \\
\textbf{e}_{\cdot x} &= [0, 1, 0, 0, 1, 1, 0, -1, 0, 0, 1, 1, -1, -1, 0, 0, 1, -1, -1, 0, -1, 1, 1, 1, -1, -1, -1]^\top, \\
\textbf{e}_{\cdot y} &= [0, 0, 1, 0, 1, 0, 1, 0, -1, 0, -1, 0, 1, 0, 1, -1, 1, -1, 0, -1, 1, -1, 1, -1, 1, -1, -1]^\top, \\
\textbf{e}_{\cdot z} &= [0, 0, 0, 1, 0, 1, 1, 0, 0, -1, 0, -1, 0, 1, -1, 1, 1, 0, -1, -1, 1, 1, -1, -1, -1, 1, -1]^\top.
\end{align}\end{linenomath}
In general, a different set of velocity vectors $\mathbf{e}$ can be chosen for the evolution of $\mathbf{f}$ and $\mathbf{h}$.

As many equations are formulated analogously for both $\textbf{f}$ and $\textbf{h}$, the letter $\textbf{g}$ will be used as placeholder for both of these distribution functions.
In this study, the collision operator $\Omega$ is considered to consist of three steps: 
some transformation $\mathscr{W}$, relaxation to equilibrium, and inverse of the transformation,
\begin{linenomath}\begin{align}
\Omega_\text{G}(\mathbf{g}) = \mathscr{W}^{-1}\Big(\mathscr{W}(\mathbf{g}) + \mathbb{S}\big(\mathscr{W}(\mathbf{g}^\eq) - \mathscr{W}(\mathbf{g})\big)\Big),
\end{align}\end{linenomath}
where $\mathbb{S}$ is the relaxation matrix.
The operator $\mathscr{W}$ can be the transformation to the space of {\it moments}, {\it central moments} or {\it cumulants}.

The moments of distribution function can be calculated in a stationary or a moving reference frame~\cite{Geier2006}.
The so-called {\it cascaded} or central moments are moments calculated in a reference frame, moving with the local macroscopic velocity, $\boldsymbol{u}$.
The discrete, raw and central moments are defined as,
\begin{linenomath}\begin{align}
\Upsilon^G_{mno} &= \sum_{\alpha}(e_{\alpha x})^m (e_{\alpha y})^n ( e_{\alpha z})^o g_{\alpha}, \label{eq:raw_mom_def} \\
\tilde{\Upsilon}^G_{mno} &= \sum_{\alpha} ( e_{\alpha x} - u_x)^m ( e_{\alpha y} - u_y)^n ( e_{\alpha z} - u_z)^o g_{\alpha}, \label{eq:cm_mom_def}
\end{align}\end{linenomath}
where $mno$ is a three-index, spanning triplets of positive numbers.
Notice, that for a fixed set of three-indexes $mno$, the \cref{eq:raw_mom_def,eq:cm_mom_def} forming the raw, $\mathscr{M}$, 
and central moment, $\tilde{\mathscr{M}}$, transforms can be expressed as matrix multiplication~\cite{Asinari2008,Fei2017,Fei2018b},
\begin{linenomath}\begin{align}
\boldsymbol{\Upsilon}^G =\mathscr{M}(\mathbf{g})&= \mathbb{M} \boldsymbol{g}, \label{eq:to_raw_mom} \\
\boldsymbol{\tilde{\Upsilon}}^G=\tilde{\mathscr{M}}(\mathbf{g}) &= \mathbb{N} \boldsymbol{\Upsilon}^G, \label{eq:to_cmom}
\end{align}\end{linenomath}
where $\boldsymbol{\Upsilon}$ and $\boldsymbol{\tilde{\Upsilon}}$ denote the vector containing raw and central moments, respectively.
Here, we select all indexes $mno$ for which $m$, $n$ and $o$ are less than 3, in the following order: [ {\footnotesize
000, 
100,
010, 
001, 
110, 
101,
011, 
200, 
020, 
002, 
120,
102, 
210, 
201, 
012, 
021, 
111,
220, 
202, 
022, 
211, 
121, 
112, 
122, 
212, 
221, 
222}
].
This choice of 27 moments results in non-singular square matrices $\mathbb{M}$ and $\mathbb{N}$, making the $\mathscr{M}$ and $\tilde{\mathscr{M}}$ operators reversible.
Remarks regarding assembling of the matrices can be found in \ref{app:NM_matrices}.

The density and internal energy correspond to the zeroth order moments of hydrodynamic and energy distribution functions respectively.
The momentum $\rho\mathbf{u}$ corresponds to the first moment of $\mathbf{f}$.
According to the probabilistic definition, the cumulants or central moments of some quantity $\mathbf{g}$ shall be calculated basing on $\mathbf{g}$ itself.
Observe, that the {\it cascaded} or central moments, defined here,
would correspond to the probabilistic definition of central moments in case hydrodynamics distributions, $\mathbf{f}$, but do not for the advected ones, $\mathbf{h}$.
Strictly speaking, the macroscopic velocity, $\mathbf{u}$, being used to compute the shift matrix, $\mathbb{N}=\mathbb{N}(\mathbf{u})$, is related to the first order moment of $\mathbf{f}$ not $\mathbf{h}$.

A detail presentation of the cumulant transformation, $\mathscr{C}$, and the use of cumulants in \ac{LBM}, can be found in the works of Geier~\cite{Geier2015a,Geier2017a} and Coreixas~\cite{Coreixas2019,Coreixas2020}.
In this work, the cumulant transform for both the advected and hydrodynamic field is the same and follows the rules described in~\cite{Geier2015a}.
Generally, one can think of cumulants as of intensive and statistically independent quantities as opposed to (central) moments, being extensive ones.
In case of the hydrodynamic field, the first order moment correspond to momentum, $\rho \boldsymbol{u}$, as opposed to first order cumulant which is just the velocity, $\boldsymbol{u}$. 

\subsubsection{Considered collision kernels}
In the following subsection, four different collision kernels being investigated in the present study will be described.
In all cases, the hydrodynamics collision operator $\Omega_\text{F}$ is always based on cumulant transform $\mathscr{C}$ and follows the formulas described in~\cite{Geier2015a}.
Denoting cumulants as $\boldsymbol{\mathcal{C}}^\text{G}=\mathscr{C}(\mathbf{g})$, the cumulant collision kernel can be written as,
\begin{linenomath}\begin{align}
\Omega_\text{F}(\mathbf{f}) &= \mathscr{C}^{-1}\left(\boldsymbol{\mathcal{C}}^\text{F} + \mathbb{S}^\text{F} \left(\boldsymbol{\mathcal{C}}^\text{F,eq} - \boldsymbol{\mathcal{C}}^\text{F}\right)\right)\label{eq:hydro_cu_collision} 
\end{align}\end{linenomath}
On the other hand, different kernels are used for the internal energy field, based on either central moments or cumulants,
\begin{linenomath}\begin{align}
% cu internal energy field
\Omega_\text{H}(\mathbf{h}) &= \mathbb{M}^{-1} \mathbb{N}^{-1} \left(\boldsymbol{\tilde{\Upsilon}}^{\text{H}} + \mathbb{S}^\text{H} \left(\boldsymbol{\tilde{\Upsilon}}^\text{H,eq} - \boldsymbol{\tilde{\Upsilon}}^\text{H}\right)\right), \label{eq:H_cm_collision}\\
\text{or} \nonumber \\
\Omega_\text{H}(\mathbf{h}) &= \mathscr{C}^{-1}\left(\boldsymbol{\mathcal{C}}^\text{H} + \mathbb{S}^{\text{H}} \left(\boldsymbol{\mathcal{C}}^{\text{H,eq}} - \boldsymbol{\mathcal{C}}^{\text{H}} \right)\right),\label{eq:H_cu_collision}
\end{align}\end{linenomath}
The analytical form of the Maxwell-Boltzmann equilibrium distribution is used to calculate the central moments for the internal energy field (see \ref{app:edf_discretization}).
The non-zero elements of equilibrium vector for $\mathbf{h}^{eq}$ are,
\begin{linenomath}\begin{align}
\boldsymbol{\tilde{\Upsilon}}^{\text{H,eq}} 
=
& \left[
 \Upsilon_{000}^{eq}, 
 ...0...,	              \hspace{0.2em}
 \Upsilon_{200}^{eq}, \hspace{0.2em}
 \Upsilon_{020}^{eq}, \hspace{0.2em}
 \Upsilon_{002}^{eq}, \hspace{0.2em}
 ...0...,                 \hspace{0.2em}
 \Upsilon_{220}^{eq}, \hspace{0.2em}
 \Upsilon_{202}^{eq}, \hspace{0.2em}
 \Upsilon_{022}^{eq}, \hspace{0.2em}
 ...0..., 				   \hspace{0.2em}
 \Upsilon_{222}^{eq}  %\hspace{0.2em}
 \right]^\top
 \nonumber \\
 = 
& \left[
 H, \hspace{1em}
 ...0...,	  \hspace{0.2em}
 H c_s^2, \hspace{0.4em}
 H c_s^2, \hspace{0.4em}
 H c_s^2, \hspace{0.4em}
 ...0..., 	   \hspace{0.2em}
 H c_s^4, \hspace{0.4em} 
 H c_s^4, \hspace{0.4em}
 H c_s^4, \hspace{0.4em} 
 ...0...,     \hspace{0.2em}
 H c_s^6  %\hspace{0.2em} 
 \right]^\top.
\end{align}\end{linenomath}
In case of cumulants, the collision kernel which has been originally implemented for the hydrodynamic field~\cite{Geier2015a} requires adjustments to reflect the macroscopic advection-diffusion equation.
Since the first and second-order cumulants corresponds to the mean and variance of the distribution, the equilibrium cumulants in case of an advected field read,
\begin{linenomath}\begin{align}
 \boldsymbol{\mathcal{C}}^{\text{H,eq}}
 = 
& \left[
 c_{000}^{eq}, 
 c_{100}^{eq}, 
 c_{010}^{eq}, 
 c_{001}^{eq}, 
 c_{110}^{eq},
 c_{101}^{eq},
 c_{011}^{eq},
 c_{200}^{eq},
 c_{020}^{eq},
 c_{002}^{eq},
 ...c_{ijk}^{eq}...,
 c_{122}^{eq},
 c_{212}^{eq},
 c_{221}^{eq},
 c_{222}^{eq}
 \right]^\top  \nonumber \\
 = 
& \left[
 H, \hspace{0.7em}
 u_x, \hspace{0.7em}
 u_y, \hspace{0.7em}
 u_z, \hspace{0.7em}
 0, \hspace{1.1em}
 0, \hspace{1.1em}
 0, \hspace{1.1em} 
 c_s^2, \hspace{0.7em}
 c_s^2, \hspace{0.7em}
 c_s^2, \hspace{0.7em}
  ...0..., \hspace{0.9em} 
 0, \hspace{1.0em} 
 0, \hspace{1.0em}
 0, \hspace{1.0em} 
 0 \hspace{1.0em} 
 \right]^\top,
\end{align}\end{linenomath}
where $c_s$ refers to the lattice speed of sound and is set to $\sqrt{1/3}$.

In the present study, a diagonal relaxation matrix is used for the advection-diffusion equation and can be formulated in a general form as,
\begin{linenomath}\begin{align}
 \mathbb{S}^{\text{H}} &= \text{diag}\left([
 s_{000}, 
 s_{100}, 
 s_{010}, 
 s_{001},
 s_{110},
 ...s_{ijk}...,
 s_{222}
 ]\right),
\end{align}\end{linenomath}
where three-indexes $ijk$ correspond to the choice of moment three-indexes.
The main relaxation frequency, $s^H$, corresponds to the macroscopic thermal conductivity~\cite{SauroSucci2019}.
As the density fluctuations are negligible ($\ll 0.01 \%$) for the cases being investigated, the main relaxation frequency can be expressed as~\cite{Shi2004}:
\begin{linenomath}\begin{align}
 s^H &= \frac{1}{\frac{k}{c_s^2 \delta t} + 1/2}.
\end{align}\end{linenomath}

Based on this framework, four collision operators are formulated.
Three of them use the central moment transform and different choices of relaxation frequencies $s_{ijk}$, and one uses the cumulant transform.

\textbf{CM-SRT} is obtained by setting $ s_{ijk} = s^H$.
Thanks to  $ \mathbb{S}^{\text{H}} $ being diagonal, the \ac{CM}-\ac{SRT} relaxation scheme is equivalent to the well known \acf{SRT} scheme in the space of distribution functions,
\begin{linenomath}\begin{align}
\boldsymbol{h}^{\star}(\textbf{x}, t)
&= \mathbb{M}^{-1} \mathbb{N}^{-1} \boldsymbol{\tilde{\Upsilon}}^{\text{H},\star}(\textbf{x}, t)  \nonumber \\
&= \mathbb{M}^{-1} \mathbb{N}^{-1} 
\left[ (\mathbbm{1} - \mathbb{S}^{\text{H}})
\mathbb{N} \mathbb{M}  \boldsymbol{h} + \mathbb{S}^{\text{H}} \boldsymbol{\tilde{\Upsilon}}^{\text{H, eq}}
\right]  \nonumber \\
&= (1-s^H)\boldsymbol{h} 
+ \underbrace{\mathbbm{1} s^H}_{\mathbb{S}^{\text{H}} } 
\underbrace{\mathbb{M}^{-1} \mathbb{N}^{-1} \boldsymbol{\tilde{\Upsilon}}^{\text{H, eq}}}_{\boldsymbol{h}^{eq}} \nonumber \\
&= (1-s^H)\boldsymbol{h} + s^H \boldsymbol{h}^{eq}.
\end{align}\end{linenomath}

\textbf{CM-$1^{st}$} is the basic approach for (central) moment based scheme for advection diffusion equation. 
Only the first order moments are relaxed ($ s_{ijk} =  s^H$ for $i+j+k = 1$),
while the higher order central moments are set to equilibrium ($ s_{ijk} = 1$ for $i+j+k > 1$).
Similar relaxation approach has been adopted in~\cite{Hajabdollahi2018,Elseid2018}, 
although it has been presented in a generalized framework, indicating that the remaining relaxation rates can be tuned independently to influence numerical stability.
Benchmarks conducted in the current study confirm, that the relaxation rates responsible for the higher order moments must be adjusted to mitigate \textit{wiggles} occurring at numerically low conductivities. 

\textbf{CM-TRT} origins from the \acf{TRT} scheme which has been derived by Ginzburg in 2005~\cite{Ginzburg2005,Ginzburg2005a},
and extended in the subsequent years~\cite{Ginzburg2008,Servan-Camas2008,Ginzburg2010,Kuzmin2011b}.
The basic idea is to separate the relaxation rate of the odd and even moments. 
Usually, the specific combination of the relaxation rates, known as a \textit{magic parameter} is kept constant.
As a result, the stationary, non-dimensional solution of \ac{NS} or \ac{ADE} is exactly controlled by the similarity numbers~\cite{Kuzmin2011b}.
One of the beneficial consequences is that the transport coefficients (like viscosity, conductivity) shall not influence the apparent location of the boundary condition~\cite{Ginzburg2008,Cui2016}.
It has been shown in~\cite{Huang2015a,Lu2019} that the TRT allows to eliminate an unphysical numerical diffusion in solid-liquid phase change model.
However, the authors of the current study decided to not set the \textit{magic parameter} for the following reasons:
\setlist{nolistsep}
\begin{enumerate}[noitemsep,label=(\alph*)]
\item The central moments' relaxation scheme does not collapse to the \ac{TRT} defined in the space of distribution functions (see \ref{app:TRT}).
\item The second order boundary conditions based on the linear interpolation schemes~\cite{HamedBouzidi2001,Dubois2019} (see \cref{sec:BC}), which have been used in the current study do not preserve the effect of the magic parameter~\cite{Pan2006,Khirevich2015}. 
According to~\cite{DHumieres2009,Ginzburg2008}, application of \textit{magic} boundary schemes allows to obtain viscosity independent permeability, however such extension is beyond the scope of the present work.
\item There is no universal, most accurate magic number~\cite{Ginzburg2009viscosity}.
\end{enumerate} 
In this contribution, the odd-moments are relaxed with a common rate ($s_{odd} = s^H$), while the even moments are set to equilibrium ($s_{even} = 1$).
For further discussion regarding the relation with the original \ac{TRT} model, the interested reader is referred to \ref{app:TRT}.

\textbf{Cumulants-$1^{st}$} follows the statistical independence of cumulants, 
thus only the first order cumulants are relaxed with $s_{ijk} =  s^H$ (for $i+j+k = 1$),
while the higher order ones are set to equilibrium values ($ s_{ijk} = 1$ for $i+j+k > 1$).

\section{Boundary Conditions (BC)} \label{sec:BC}
This section briefly describes the boundary conditions used within the current work.
The simple \ac{BB} rule reduces the convergence of the \ac{LBM} to first order if the wall is not located exactly between lattice nodes.
The higher-order \acl{BC} are implemented in the present study to circumvent this issue and recover the second order convergence.

\begin{figure}[htbp]
\begin{center}
   \includegraphics[width=0.6 \textwidth]{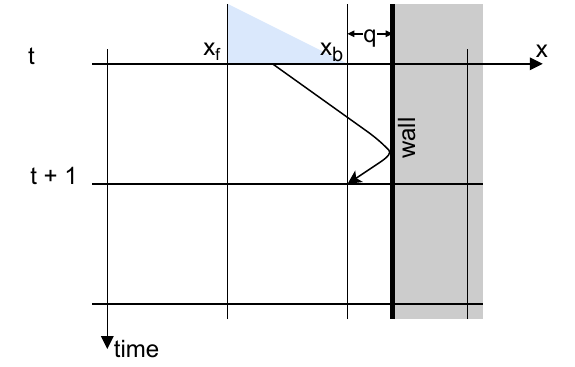}  
   \caption{Interpolated (Anti-)Bounce Back in the time-space diagram.
   The blue triangle indicates the intensity of the interpolation function 
   between $g_i^{\star}(\boldsymbol{x}_b,t)$ and $g_i^{\star}(\boldsymbol{x}_f,t)$ for $q \in \left[0, 0.5 \right]$. }
   \label{fig:interpolatedBB}
 \end{center}
\end{figure}

~\citet{HamedBouzidi2001} proposed the \acf{IBB} scheme to represent a non-slip wall on a curved boundary.
It is assumed that during each streaming step, the distributions travels a distance $|\boldsymbol{e}_i| \Delta t$.
The walls are modelled by a bounce back of said distributions back into the domain.
As the wall is not necessarily half-way between lattice nodes, the algorithm employs a linear interpolation scheme between boundary node $\boldsymbol{x}_b$ and neighbouring fluid node $\boldsymbol{x}_f$,
\begin{linenomath}\begin{align}  
f_{\bar{\alpha}}(\boldsymbol{x}_b,t + \Delta t)  &=
\left\{ 
	\begin{array}{ll}
				2q f_\alpha^{\star}(\boldsymbol{x}_b,t) + (1-2q)f_\alpha^{\star}(\boldsymbol{x}_f,t)
			  \hspace{4.5em} \text{for} \, q \in \left[0, 0.5 \right]  \vspace{1.5em} \\
				\frac{1}{2q} \left[ f_\alpha^{\star}(\boldsymbol{x}_b,t) + (2q-1) f_{\bar{\alpha}}^{\star}(\boldsymbol{x}_b,t) \right]
			 \hspace{3.5em} \text{for} \,q \in \left( 0.5, 1 \right],
	\end{array} 
\right. \label{eq:IBB}
\end{align}\end{linenomath}
where $q$ is the distance (along lattice link) between the boundary node and the actual boundary and $\bar{\alpha}$ denotes direction opposite to $\alpha$.
The idea is depicted in \cref{fig:interpolatedBB}.

The interpolation scheme proposed by~\citet{HamedBouzidi2001} can be modified~\cite{Dubois2009,Ginzburg2009viscosity,Li2013b,Dubois2019} to obtain Dirichlet boundary conditions for advection-diffusion problems.
Due to change in sign of the post-collision distribution function, the scheme is refereed as \acf{IABB}, 
\begin{linenomath}\begin{align}  
h_{\bar{i}}(\boldsymbol{x}_b,t + \Delta t)  &=
\left\{ 
	\begin{array}{ll}
		- \left[ 2q h_i^{\star}(\boldsymbol{x}_b,t) + (1-2q)h_i^{\star}(\boldsymbol{x}_f,t) \right] + 2 h_i^{eq}(\boldsymbol{x}_w,\boldsymbol{u},t)
		\hspace{3.0em} \text{for} \, q \in \left[0, 0.5 \right]  \vspace{1.5em} \\
		\frac{1}{2q} \left[ - h_i^{\star}(\boldsymbol{x}_b,t)  + (2q-1) h_{\bar{i}}^{\star}(\boldsymbol{x}_b,t)  + 2 h_i^{eq}(\boldsymbol{x}_w,\boldsymbol{u},t) \right]
			 \hspace{3.0em} \text{for} \,q \in \left( 0.5, 1 \right],
	\end{array} 
\right. \label{eq:IABB}
\end{align}\end{linenomath}
where $h_i^{eq}(\boldsymbol{x}_w,\boldsymbol{u},t)$ is a source term designed to impose the desired temperature at the wall,~$\boldsymbol{x}_w$.
If the interpolation is skipped, then the scheme simplifies to the so called \ac{ABB}~\cite{Ginzburg2008,Ginzburg2008b,Izquierdo2008}.
In the simplest form of the Dirichlet boundary condition, known as the \ac{EQ}~\cite{He1997_nonslipbc,Latt2008,Mohamad2009}, only the equilibrium part of the above equation is prescribed to the outgoing distributions.

\section{Model Verification and Validation}\label{sec:benchmarks}
In this section, a set of numerical experiments used to evaluate the accuracy of the investigated kernels is described.
The benchmarks starts with advection-diffusion problems in a fixed, external velocity field.
Next, the second order boundary conditions are tested.
Finally, a comprehensive study of a flow with forced convection is performed.

To assure consistency all cases are calculated with the same 3D code. 
In the 2D cases, the z-direction is periodic and it is cut to three elements. 
This is technical minimum in our solver~\cite{Laniewski-Wollk2016a,TCLB}, because a message passing layer is used for domain decomposition. 
In such a setup, the 3D model reduce itself to 2D one (of course, with a higher computational cost).
Observe, that the LBM weights of D3Q27 lattice, when summed over the z-direction, give the standard LBM weights for D2Q9.

The normalized $L_2$ norm of error is utilised to compare the results between the simulations.
It is defined as,
\begin{linenomath}\begin{align}
L_2 =  \sqrt{ \dfrac{\sum_i\left(T^{analytical}_i - T^{numerical}_i\right)^2}{\sum_i \left(T^{analytical}_i \right)^2}}
\end{align}\end{linenomath}
where the sum $\sum_i$ goes over all lattice nodes and  $T_i$ is the temperature in the $i-th$ node.

\subsection{Advection-Diffusion of a Gaussian hill}
To avoid the influence of boundary conditions, the first benchmarks investigates behaviour of the collision kernels in a periodic domain.
In the case of an isotropic diffusion, and convection with constant velocity, the analytical solution describing the evolution of a Gaussian hill can be derived~\cite{ginzburg2005equilibrium,Yoshida2010,Kruger2017}.
The formula can be expressed as,
\begin{linenomath}\begin{align}
C(\boldsymbol{x}, t)=\frac{\left(2\pi\sigma_{0}^{2}\right)^{N/2} }{\left(2\pi(\sigma_{0}^{2} + 2 k t)\right)^{N/2}} 
C_0 \exp \left(-\frac{\left(\boldsymbol{x}-\boldsymbol{x}_{0}-\boldsymbol{u} t\right)^{2}}{2\left(\sigma_{0}^{2}+ 2 k t\right)}\right) \label{eq:GaussianHill_anal_solution}
\end{align}\end{linenomath}
where $N$ is the number of spatial dimensions, $t$ is time and $\sigma_{0}$ represents the initial variance of the distribution.

\begin{figure}[htbp]
        \centering
        \begin{subfigure}[b]{0.485\textwidth}
            \centering
        \includegraphics[width=\linewidth]{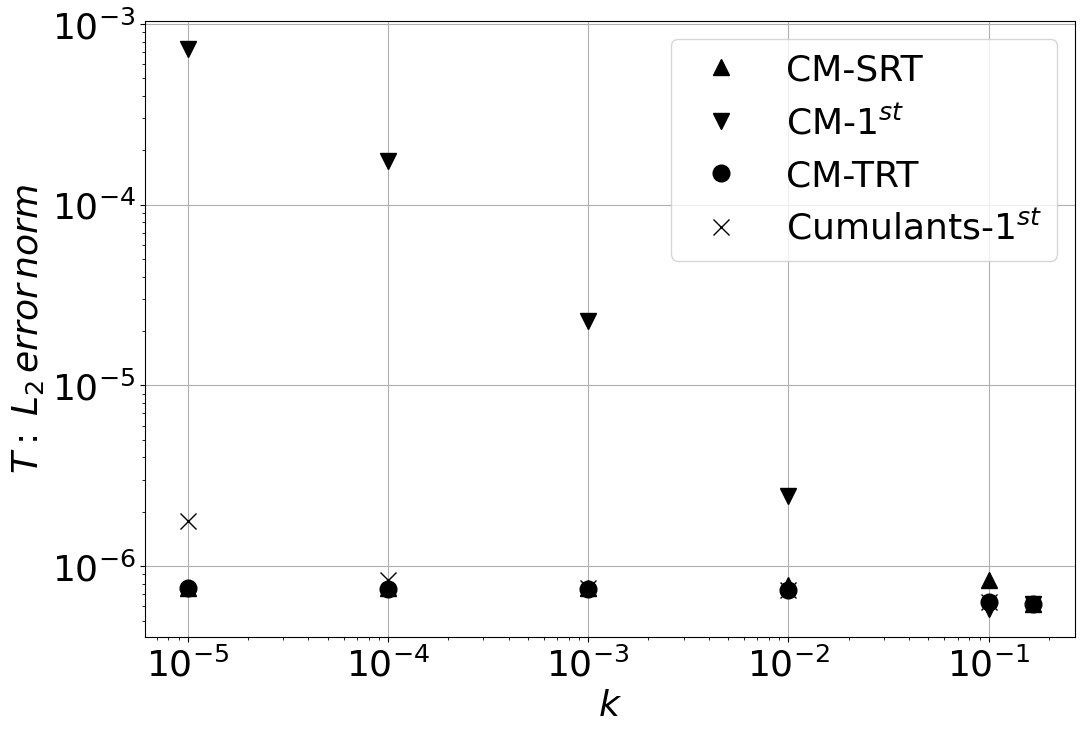}%
            \caption{Without external velocity field, $u_x=0$} 
            \label{fig:GaussiaHill2D_ux0_convergence}
        \end{subfigure}
        \hfill
        \begin{subfigure}[b]{0.485\textwidth}  
            \centering 
        \includegraphics[width=\linewidth]{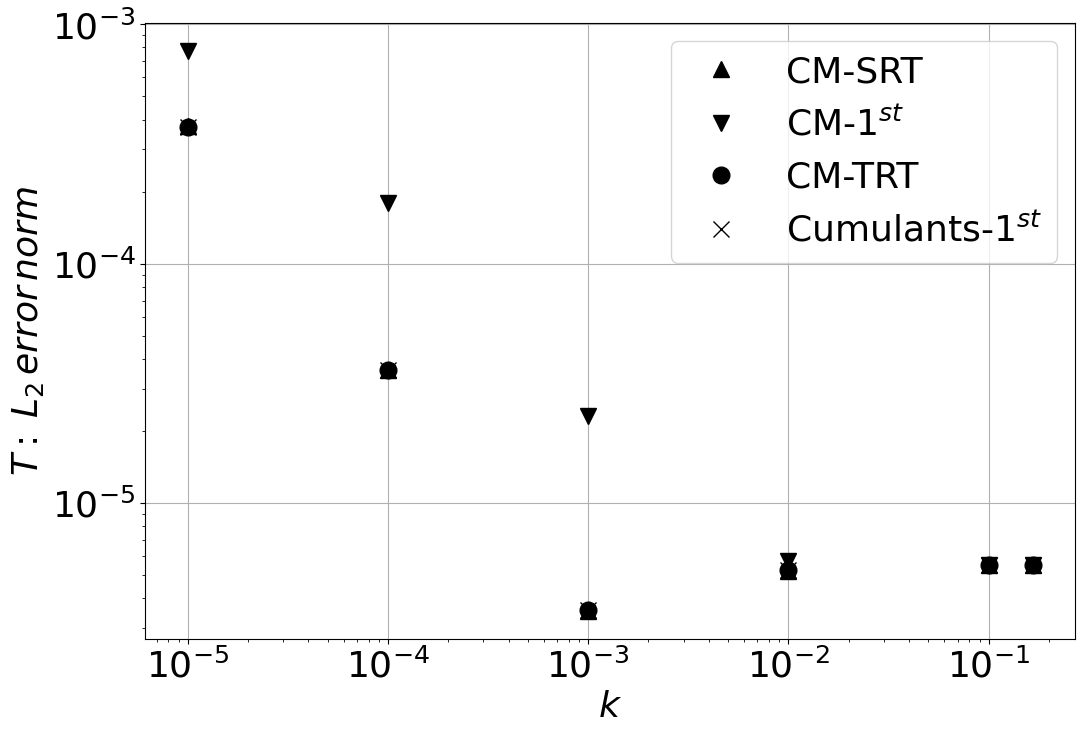}
            \caption{External velocity field, $u_x=0.1$}   
            \label{fig:GaussiaHill2D_ux01_convergence}
        \end{subfigure}
    \caption{Advection-diffusion of a Gaussian hill has been calculated analytically for the physical time $t_{SI}=100$ and physical conductivity $k_{SI} =4 $. 
    To benchmark the investigated collision kernels, the same physical case has been simulated using different numerical conductivities $k$ on a $256 \times 256 \times 3$ lattice.}
	\label{fig:GaussianHill2D}
    \end{figure}

Given the initial condition, the same physical case (defined by physical time, $t_{SI}$, physical length, $L_{SI}$, and physical conductivity $k_{SI}$), can be simulated numerically using different time steps $\delta t= (\delta x)^2 k_{LB}/k_{SI}$. 
The $ \delta x$ denotes ratio of the physical length to the number of lattice nodes and has been fixed as $ \delta x = L_{SI}/L = 1$. 
Since $t_{SI} = n \delta t$, where $n$ is the number of iterations, the $k_{LB}$ can be expressed as $k_{LB} = (t_{SI} k_{SI})/ \left(n (\delta x)^2 \right)$.
The domain was a square, $256 \times 256 \times 3$, with periodic boundaries.
Each case has been initialised with a 2D Gaussian distribution, according to \cref{eq:GaussianHill_anal_solution}, with initial variance, $\sigma_0=100$.
The simulations were executed for $n$ iterations, ranging from \num{2400} (corresponding to $k_{LB}=1/6$) to \num{40e6} ($k_{LB}=10^{-5}$).
Once the simulation reached the prescribed number of iterations, the result was compared against the analytical solution. 
The $L_2$ error norm has been plotted in \cref{fig:GaussianHill2D} for each of the investigated kernels.
For pure diffusion, the error of all collision kernels is small and of similar level, for $k$ ranging from $1/6$ to $10^{-4}$ (see \Cref{fig:GaussiaHill2D_ux0_convergence}), except for CM-$1^{st}$ model, for which the error strongly depends on $k$.
In the case of constant velocity $\mathbf{u}=[1,0]$, the error of all collision kernels depend on $k$, 
as artefacts resulting from numerical dispersion dominates the ones related to diffusion (see~\Cref{fig:GaussiaHill2D_ux01_convergence}).
The geometrical structure of errors ({\it wiggles}) caused by the numerical dispersion is shown in~\cref{sec:sq}.

To show the behaviour of the investigated kernels in 3D, and the geometry of the introduced error, a spherical Gaussian hill was investigated on a \DQQ{3}{27}{27} lattice.
The domain has initialised according to \cref{eq:GaussianHill_anal_solution} with initial variance of $\sigma_{0}^{2} = 100$.
The domain was a box, $256 \times 256 \times 256$, with periodic boundary conditions.
Conductivity of the stationary medium has been set to moderate value, namely $k_{LB} = 10^{-3}$ and the simulation was run for $n = 400 000$ iterations.
   \begin{figure}[htbp]
        \centering
        \includegraphics[width=0.7\textwidth]{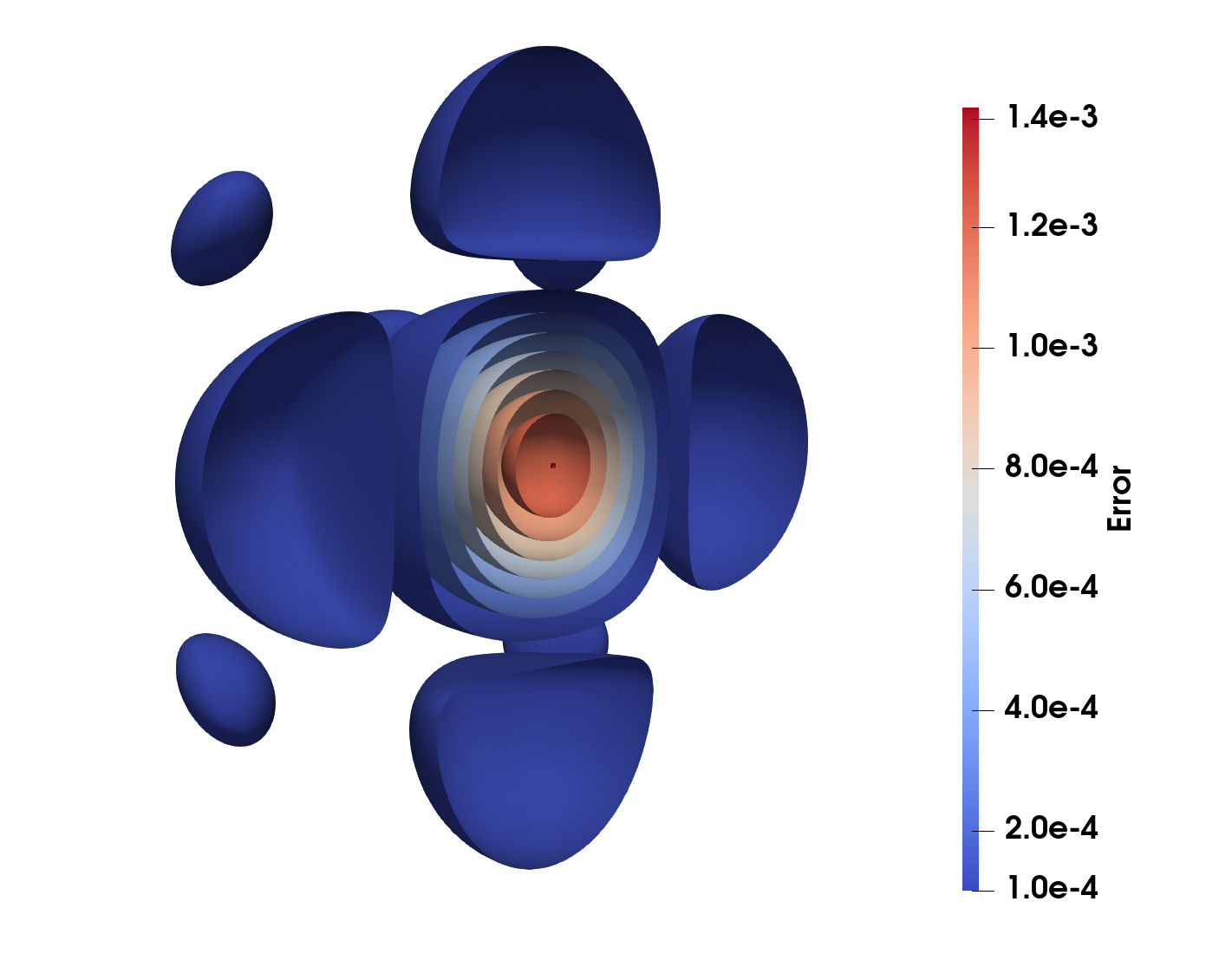}
        \caption{The iso-contours of the absolute error of the 3D Gaussian hill for the \ac{CM}-$1^{st}$ kernel.
        The simulation has been run for $400 000$ iterations with conductivity being set to $k_{LB} = 10^{-3}$.} 
\label{fig:GaussianSphere}
\end{figure}
Again, the results were compared with analytical solution.
All the collision kernels resulted in error with spherical symmetry, except for the \ac{CM}-$1^{st}$ kernel.
The geometrical structure of the absolute error for this collision operator is shown \cref{fig:GaussianSphere}.
It is immediately evident that when only first-order central moments are relaxed, a considerable amount of spurious, mesh aligned, structures arise.
Contrastingly, the other collision operators provided nearly identical results, with spherical symmetry and an order of magnitude smaller maximum errors, compared to \ac{CM}-$1^{st}$.

\subsection{Advection-diffusion of a square indicator function}\label{sec:sq}
To depict the character of wiggles caused by both external velocity field and a jump in the value of a scalar field, 
a simple advection-diffusion study with uniform velocity field and periodic boundary conditions has been done. 
The qualitative results are presented in \cref{tab:square_advection_table}.
The dimensions of the domain were $128 \times 128 \times 3$ [lu].
In the middle of the domain, a $48 \times 48 \times 3$ square has been initialised with $T = 11$, while $T_0 = 10$.
The conductivity has been set to \num{E-05}. 
Each simulation has been run for 12 800 iterations.
To store the scalar field, the \DQQ{3}{}{7} lattice has been benchmarked against \DQQ{3}{}{27}.
It is easy to observe that the \DQ{3}{7} lattice is a subset of \DQ{3}{27}.
It can be obtained by limiting the set of discrete velocities, $\textbf{e}$, to its first seven elements.
Only the first seven central moments can be represented: the zeroth, first and second order, non-diagonal ones. 
As a consequence, it is not possible to distinguish between \ac{CM}-$1^{st}$ and \ac{CM}-\ac{TRT}.
The shape of the square advected on this lattice is distorted and strong wiggles are evident.

For the \DQQ{3}{}{27} lattice, all four collision kernels were compared.
The \ac{CM}-$1^{st}$ and Cumulants-$1^{st}$ kernels generates diffusive artefacts that appears at the corners of the resting square.
The issue can be alleviated using the \ac{CM}-\ac{SRT} or \ac{CM}-\ac{TRT} kernel, which relax the higher-order moments.

%\newpage
%https://tex.stackexchange.com/questions/7208/how-to-vertically-center-the-text-of-the-cells
% \newcommand{\centered}[1]{\begin{tabular}{l} #1 \end{tabular}}
%\begin{tabular}{|l|c|c|}  <- Old version of this answer
\newcommand{\squarescale}{0.275}
%\newgeometry{top=5mm, bottom=5mm}
%\thispagestyle{empty}
%\begin{table}[htbp]\centering
\begin{table}[p]\thisfloatpagestyle{empty}\vspace{-3cm}\centering
\begin{tabular}{|@{}c@{} | @{}c@{}|@{}c@{}|@{}c@{}|}
% This update now avoids double indentations and allows hlines
\hline
% \begin{tabular}{c} \begin{sideways}\centering	\DQ{3}{7}		\end{sideways} \end{tabular} &
%  \centered{ \begin{sideways}\centering	\DQ{3}{7}		\end{sideways}	} &
  \multirow{2}{*}{\begin{sideways}\centering 	\DQ{3}{7}		\end{sideways}} &
  \centered{ \begin{sideways}\centering	\ac{CM}-\ac{SRT}		\end{sideways}	} &
  \centered{\includegraphics[width=\squarescale \linewidth]{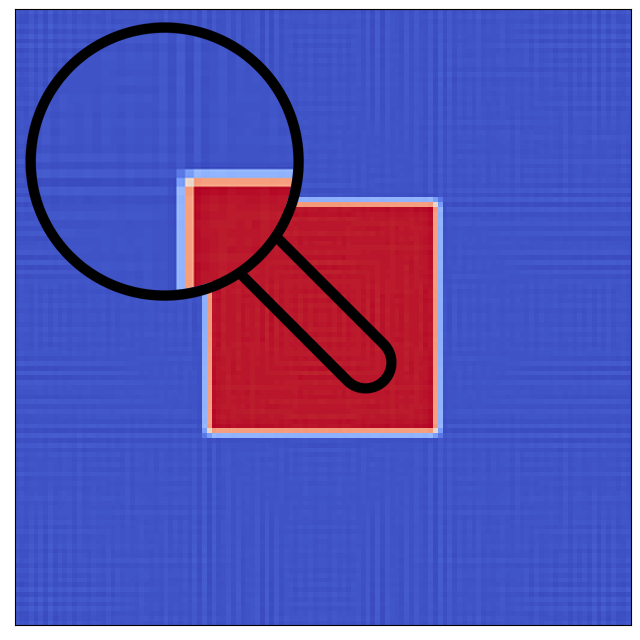}  		}&
  \centered{\includegraphics[width=\squarescale \linewidth]{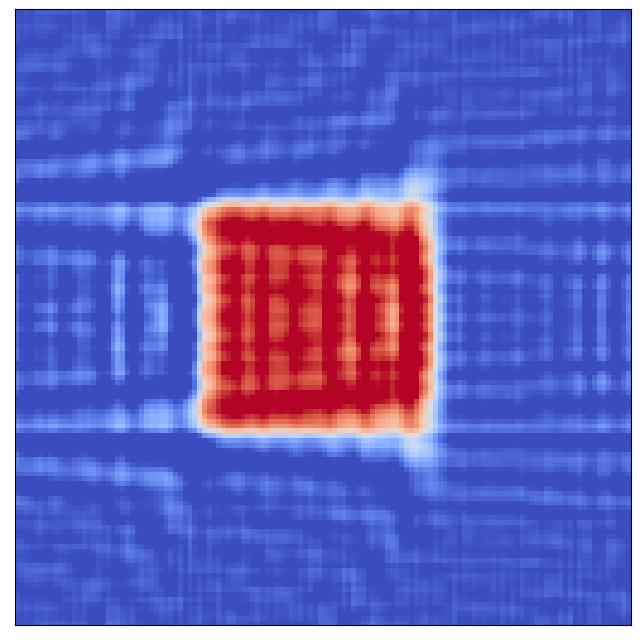}		}\\
     &
   \centered{ \begin{sideways}\centering	\ac{CM}-$1^{st}$ same as \ac{CM}-\ac{TRT}		\end{sideways}	} &
   \centered{\includegraphics[width=\squarescale \linewidth]{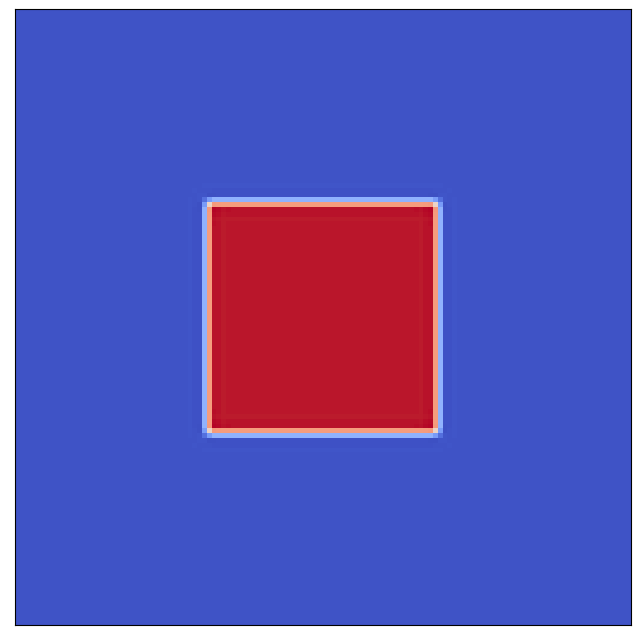}			}&
   \centered{\includegraphics[width=\squarescale \linewidth]{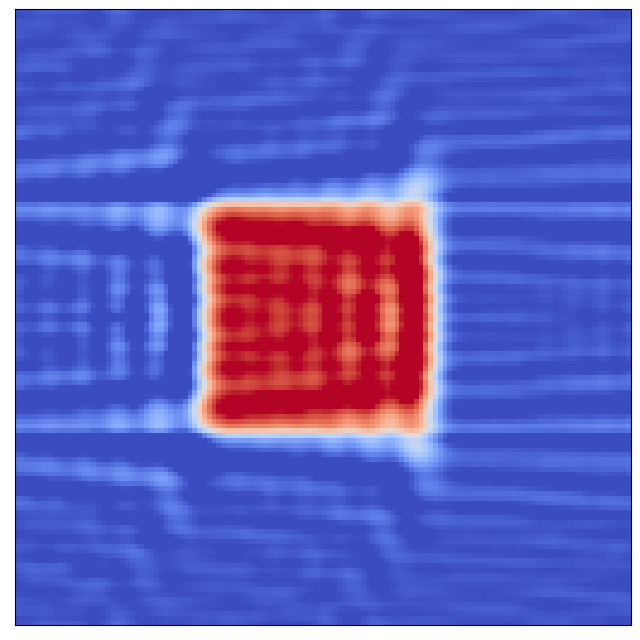}		}\\
%  \multirow{2}{*}{\begin{sideways}\centering	\DQ{3}{7}		\end{sideways}} &
%  \begin{sideways}	\centering CM-SRT		\end{sideways}	 & 
%  1 & 
%  2 \\
%    & 
%  \centering \begin{sideways}	\centering CM-SRT				\end{sideways}		 & 
%  3 & 
%  4 \\
\hline
  \multirow{4}{*}{ 	\begin{sideways} \centering	\DQ{3}{27}		\end{sideways}		} &
  \centered{ \begin{sideways}\centering	\ac{CM}-\ac{SRT}			\end{sideways}	} &
  \centered{ \includegraphics[width=\squarescale \linewidth]{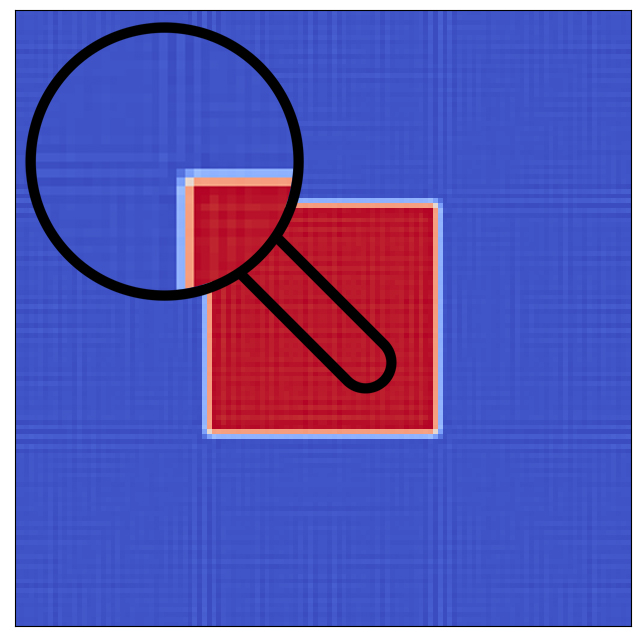}   }  &
  \centered{ \includegraphics[width=\squarescale \linewidth]{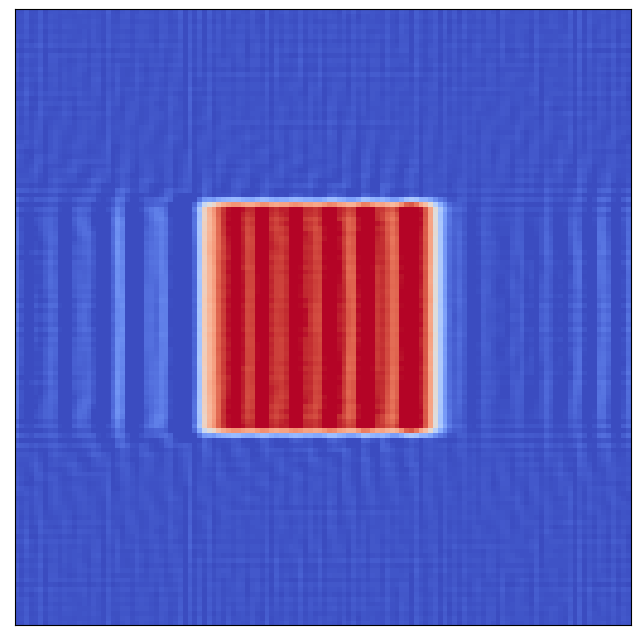}  } \\
  
  	 &
  \centered{ \begin{sideways}\centering	\ac{CM}-$1^{st}$		\end{sideways}	} &
  \centered{ \includegraphics[width=\squarescale \linewidth]{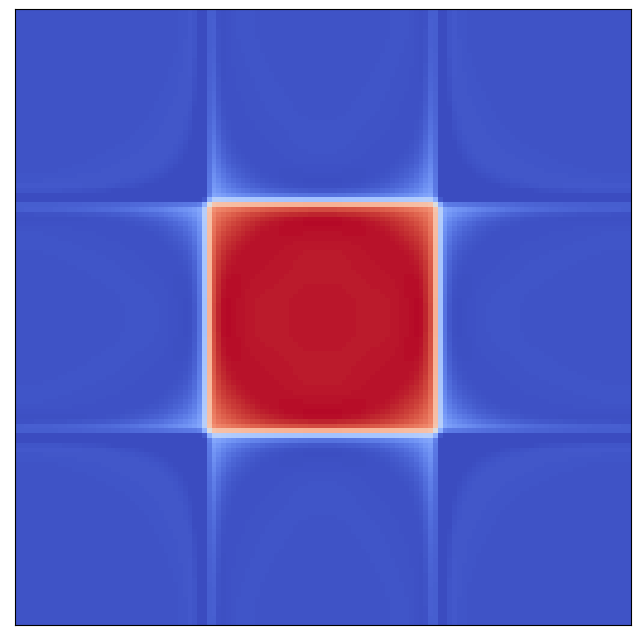}   }  &
  \centered{ \includegraphics[width=\squarescale \linewidth]{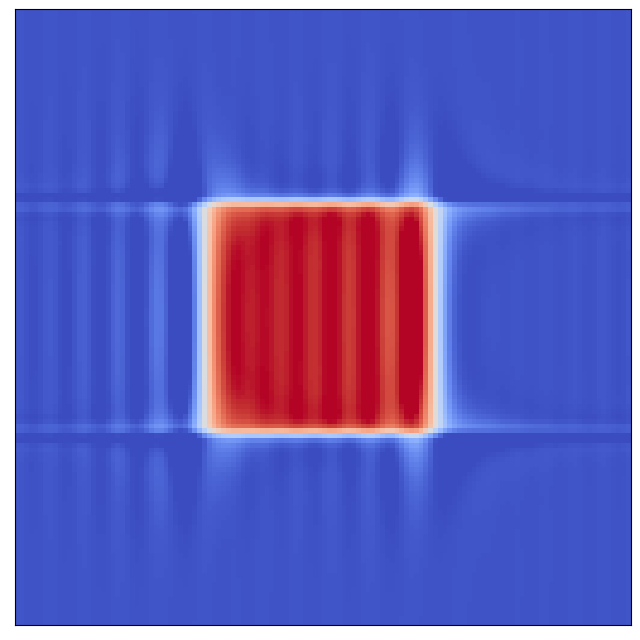}  } \\
  
  	 &
  \centered{ \begin{sideways}\centering	\ac{CM}-\ac{TRT}			\end{sideways}	} &
  \centered{ \includegraphics[width=\squarescale \linewidth]{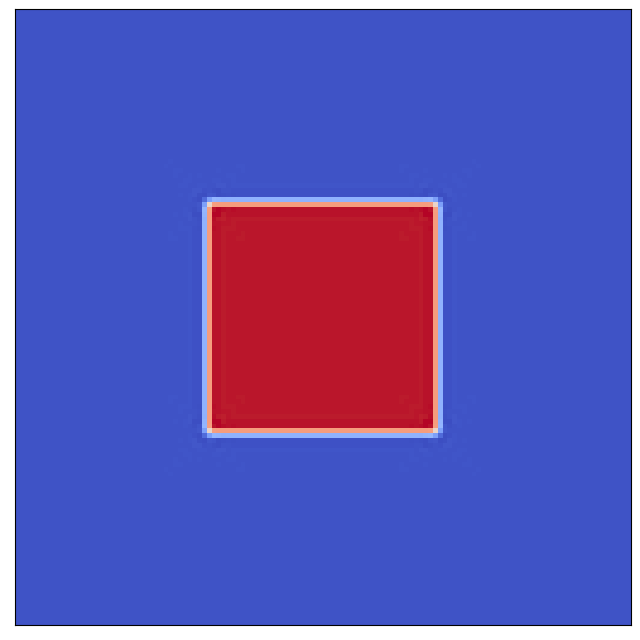}    }  &
  \centered{ \includegraphics[width=\squarescale \linewidth]{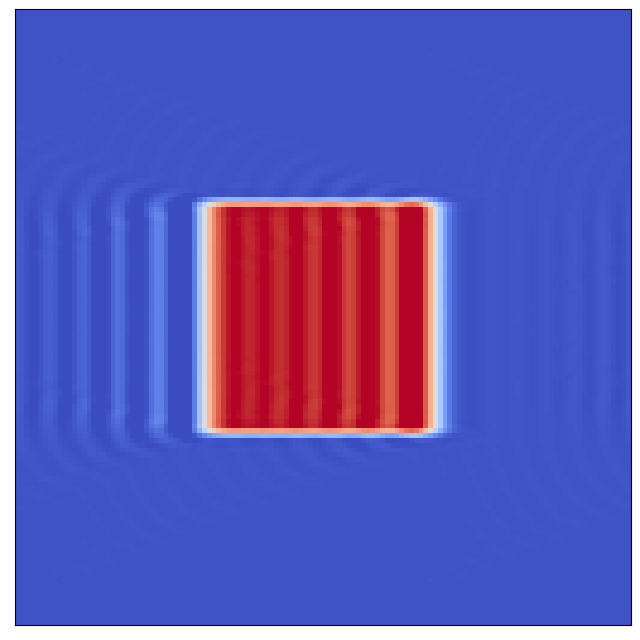}  } \\
  
  	&
  \centered{ \begin{sideways}\centering	Cumulants-$1^{st}$		\end{sideways}	} &
  \centered{ \includegraphics[width=\squarescale \linewidth]{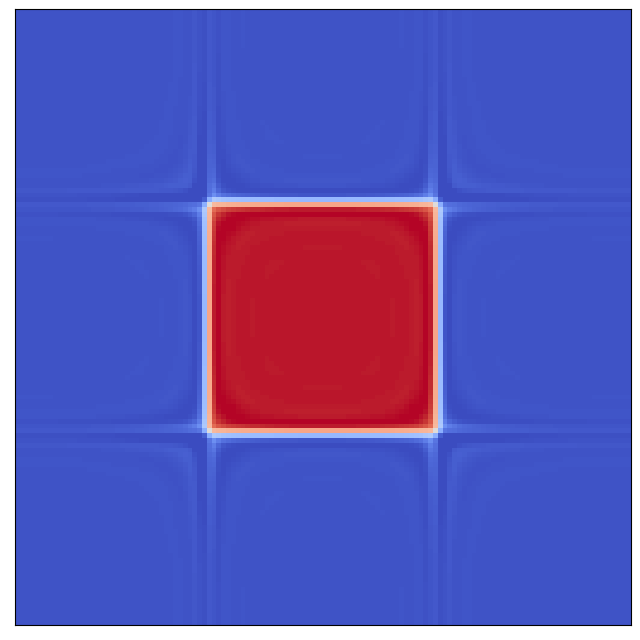}   }  &
  \centered{ \includegraphics[width=\squarescale \linewidth]{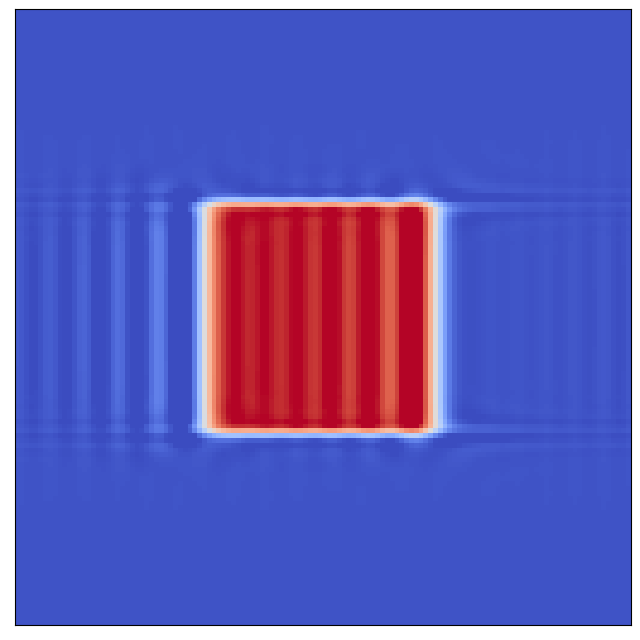}  } \\  
\hline
  \centered{ Lattice } &
  \centered{ Kernel	} &
  \centered{ $\boldsymbol{u}= \boldsymbol{0}$   }  &
  \centered{ $u_x=0.1$  } \\
\hline
\end{tabular}
\caption{
The square has been initialised with $T = 11$, while $T_0 = 10$.
The color map on all images has been clipped to range $T \in (9.98 - 11.02 )$. 
The region inside the loupe has been magnified by 175\%.
Notice the onset of numerical noise in the background when the \ac{CM}-\ac{SRT} kernel is used.
}
\label{tab:square_advection_table}
\end{table}
% \restoregeometry

\subsection{Heat conduction between two concentric cylinders}
\begin{figure}[htbp]
 	\centering
		\includegraphics[width=0.33 \textwidth]{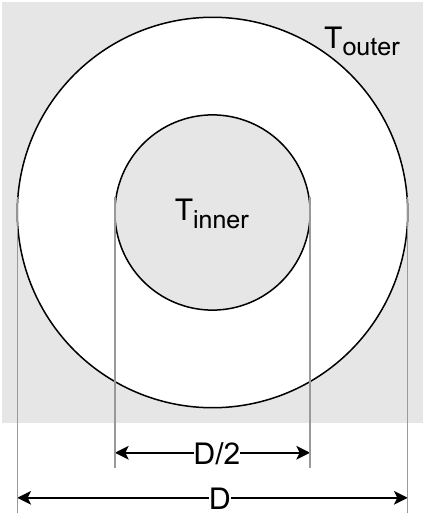} 
	\caption{ 
	The inner cylinder was a heater with diameter $\frac{1}{2}D$, while the outer one with diameter $D$ was a cooler.
	}
	\label{fig:Pipe_in_a_pipe}
\end{figure}

To asses the accuracy of a curved boundary representation, steady state heat conduction between two concentric cylinders (without flow) is studied.
The geometry is shown in \cref{fig:Pipe_in_a_pipe}.

Heat conduction in the cylindrical coordinate system can be described by a  partial differential equation,
\begin{linenomath}\begin{align}
\rho c_v \pr{t} T = \frac{1}{r}  \pr{r} \bigg(k r \pr{r} T \bigg) + \frac{1}{r^2} \pr{\phi} \bigg(k \pr{\phi} T \bigg) + \pr{z} \bigg(k \pr{z} T \bigg) + \dot{q}.
\end{align}\end{linenomath}
The general solution for a 2D, steady-state case is,
\begin{linenomath}\begin{align}
T(r) = \frac{\lambda_1}{k} ln \left( \frac{r}{r_{inner}} \right) + \lambda_2. \label{eq:general_solution_heat_conduction_on_pipe}
\end{align}\end{linenomath}
Applying the Dirichlet boundary condition for $T(r_{inner})=T_{inner}$ and $T(r_{outer})=T_{outer}$, the unknown coefficients, $\lambda_1, \lambda_2$ are find and the solution reads,
\begin{linenomath}\begin{align}
T(r) = \left(T_{outer} - T_{inner}\right)\frac{\ln{\left( \frac{r}{r_{inner}} \right)}}{\ln{\left( \frac{r_{outer}}{r_{inner}} \right)}}  + T_{inner}.
\end{align}\end{linenomath}

\begin{figure}[htbp]
    \centering
    \includegraphics[width=0.8\linewidth]{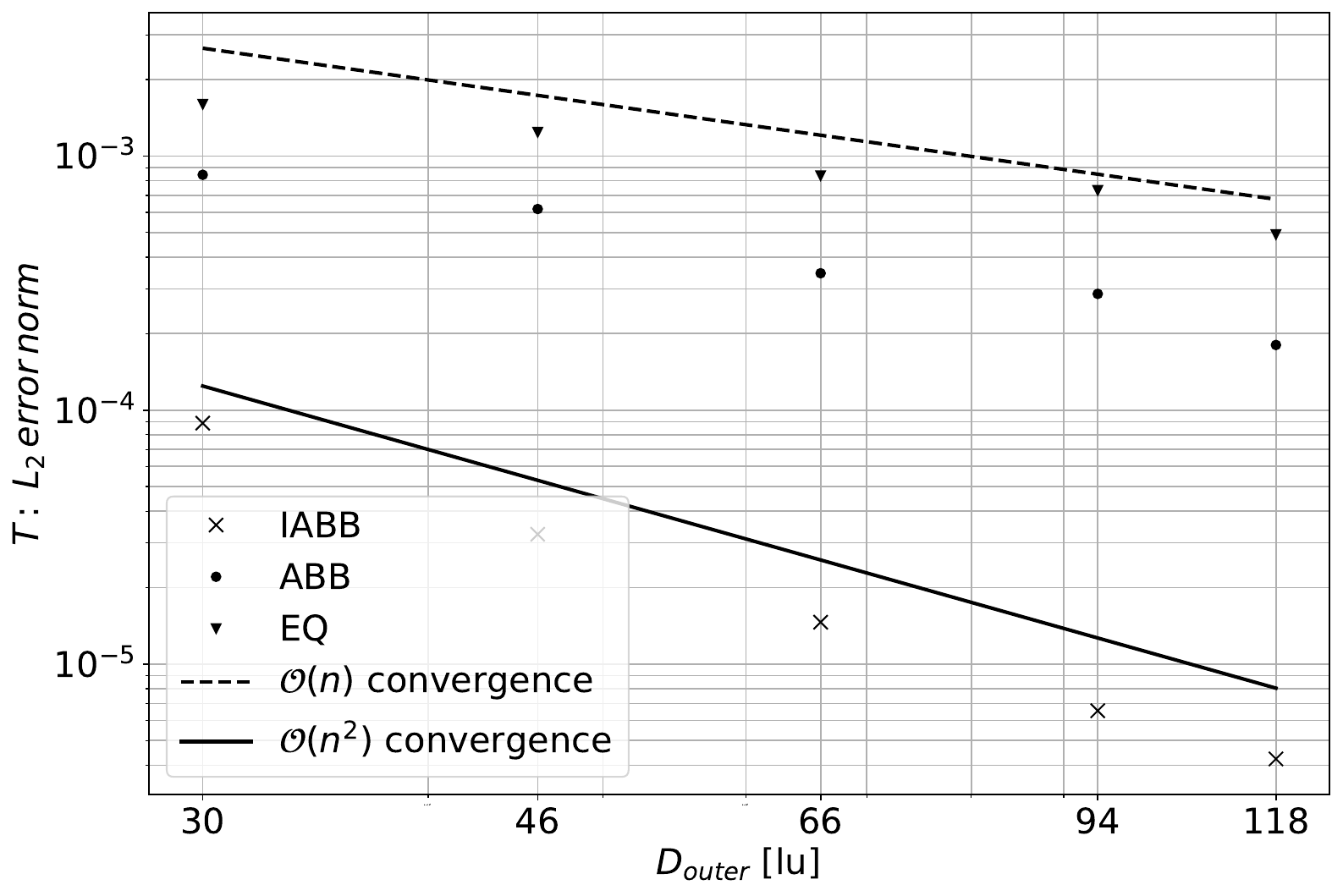}
    \caption{Grid convergence study for the Dirichlet boundary condition.
    The \ac{CM}-\ac{TRT} collision Kernel has been used and the conductivity was set to $k=0.1$. 
    Three different implementations of boundary conditions were benchmarked. The abbreviations reads:  
    \acf{IABB},
    \acf{ABB},
    \acf{EQ}.
    }
    \label{fig:IABB_vs_ABB_vs_EQ}
\end{figure}

In \cref{fig:IABB_vs_ABB_vs_EQ}, three different implementations of Dirichlet's boundary condition for the internal energy field have been assessed in a circular geometry.
As expected, only the \ac{IABB} exhibited the second order convergence rate.

\subsection{Steady, forced convective heat transfer from a confined cylinder}
\begin{figure}[htbp]
 	\centering
		\includegraphics[width=1 \textwidth]{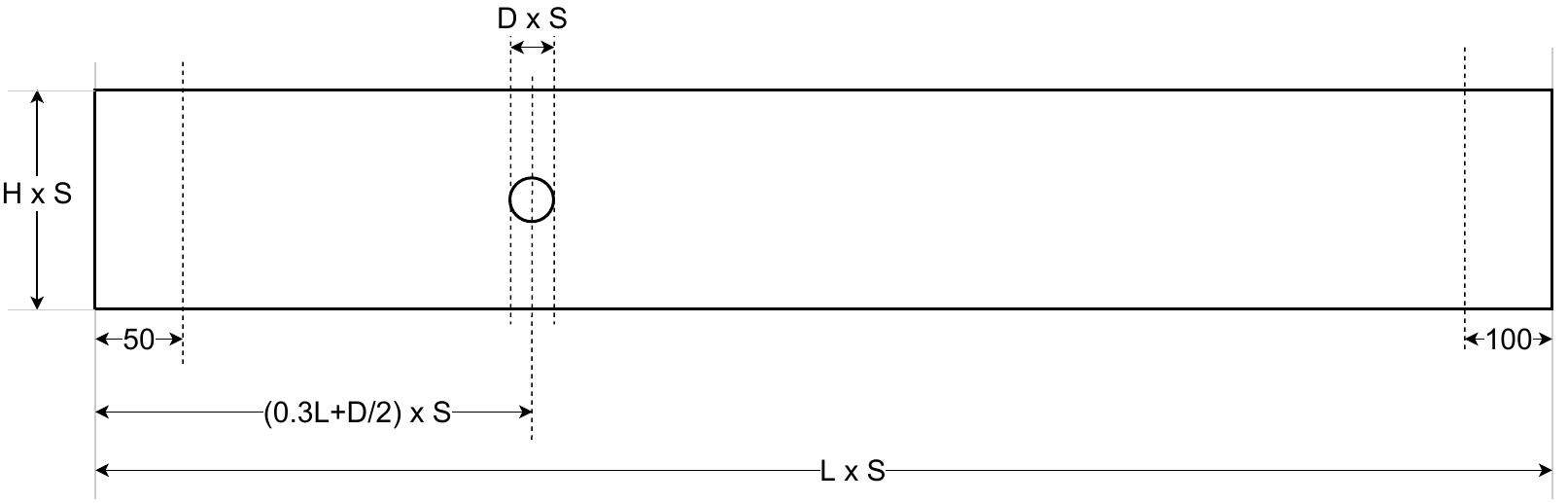} 
	\caption{Channel with hot cylinder and cross-sections used for the heat flux measurements.}
	\label{fig:HotBarmanBenchmarkGeom}
\end{figure}

From an engineering perspective, the temperature of the fluid is usually controlled by a presence of a heat exchanger.
Here, a mesh dependence study of a steady forced convection from a confined cylinder is performed to illustrate the effect of the various implementation of the boundary conditions, collision kernels at different \ac{Pr} numbers.
\cref{fig:HotBarmanBenchmarkGeom} presents parametrisation of the domain.
All simulations have been performed on a  \DQQ{3}{27}{27} lattice.
To refine the lattice, a geometry scaling factor has been defined as $S \in \left\lbrace  1,2,4 \right\rbrace $.
The height, length and cylinder's diameter of the \textit{coarse} ($S=1$) lattice are $H = 150$ [lu], $L = 1000$ [lu] and $D=30$ [lu] respectively.
On the inlet, the \textit{Zou-He}~\cite{Zou1996} \ac{BC} also know as \textit{non equlibrium bounce-back} method, with $T_{inlet}=10$, has been imposed.
The Neumann \ac{BC} has been placed at the outlet as described in~\cite{Lou2013}.
Top and bottom of the domain utilised symmetry \ac{BC}.
The Dirichlet \ac{BC} has been prescribed on walls of the cylinder, $T_{cylider}=11$, using either first or second order implementation.
In case of the first order \ac{BC}, the \acf{BB} rule for hydrodynamics and \acf{EQ} for internal energy field has been used. 
For the second order \ac{BC}, the \acf{IBB} and \acf{IABB}, as described in \cref{sec:BC} were employed.
The \ac{LBM} simulation has been iterated until heat flux calculated thorough the heater's surface, $\dot{q}_{surface}$,  matched the outflow flux, $\dot{q}_x = \dot{q}_x^{inlet} - \dot{q}_x^{outlet}$,  or the iterations limit has been reached.
The heat flux thorough the heater's surface has been calculated as,
\begin{linenomath}\begin{align}
\dot{q}_{surface} = q_{out} - q_{in} = \sum_\alpha h_\alpha^\star - \sum_\alpha h_\alpha, \label{eq:heat_flux_thorough_heater}
\end{align}\end{linenomath}
while the heat flux thorough a section as,
\begin{linenomath}\begin{align}
\dot{q}_x^{section} = \int \rho c_v T \boldsymbol{u} \cdot \boldsymbol{n} dA = \sum_\alpha e_\alpha^x h_\alpha. \label{eq:heat_flux_section}
\end{align}\end{linenomath}
To limit the effect of a boundary condition, the heat flux measurements' sections for the inlet and outlet have been defined 50 and 100 [lu] away from the boundary (see \cref{fig:HotBarmanBenchmarkGeom}).

The flow around a hot cylinder can be defined by two dimensionless numbers, namely \acf{Re} and \acf{Pr}.
The well known \ac{Re} describes the ratio of inertia to viscous forces within the fluid, $Re = \frac{u D}{\nu}$.
Subsequently, the \ac{Pr} describes the relative thickness of the momentum to thermal boundary layer, 
$Pr  = \frac{\nu}{\alpha}  = \frac{\nu \rho c_v}{k}$.
As mentioned in the \cref{sec:Intro}, it can be also viewed as a property of a medium, 
which describes the ratio of time scales at which physical phenomena related to hydrodynamic and thermal field occurs.
For fluids characterised by high $Pr$ (e.g. oil), 
the heat diffuses much slower than the momentum and the thermal boundary layer is contained within the velocity boundary layer.
In the case of liquid metals, the opposite happens. 
The \ac{Pr} is low, heat diffuses much faster than momentum, and the velocity boundary layer is fully contained within the thermal boundary layer.
Finally, the \acf{Nu} has been chosen to assess quality of the simulations.
It represents the enhancement of heat transfer through a fluid as a result of convection relative to conduction across the same fluid layer.
It is defined as $Nu=\frac{ \varsigma D}{k }$,
where $\varsigma$ is the average convective heat transfer coefficient, % $[W/(m^2 K)]$, 
$ \varsigma = \dot{q} / (A(T_{cylinder} - T_{inlet}))$ and $A=\pi D$ is the area of the cylinder. 
The heat flux, $\dot{q}$, is calculated as an average of $\dot{q}_{surface}$ and $\dot{q}_x^{section}$.

\begin{figure}[htbp]
        \centering
        \includegraphics[]{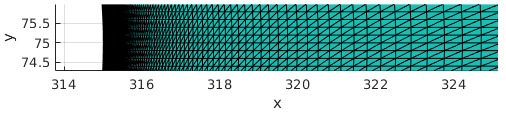}
        \caption{The FEM mesh in the vicinity of the cylinder.} 
\label{fig:FEM_mesh_vicinity_of_cylinder}
\end{figure}

In the numerical study, 72 \ac{LBM} simulations were performed. 
The resulting \ac{Nu} are compared against a high-quality solution obtained with Bubnov-Galerkin \ac{FEM} solver from QuickerSim CFD Toolbox for MATLAB.
The \ac{FEM} structural mesh consisted of 392704 second order triangular elements.
\cref{fig:FEM_mesh_vicinity_of_cylinder} presents discretization of the domain behind the cylinder.

Four collision kernels were tested with first and second order \ac{BC} for $Pr \in \left\lbrace  10,100,1000 \right\rbrace $.
To lower the computational effort, the Reynolds number was set to 10 thus, the flow pattern could be assumed to be two-dimensional. 
As a consequence, the result obtained on the \DQQ{3}{27}{27} lattice would correspond to the one obtained on \DQQ{2}{9}{9}.
The coarse lattice $1000 \times 150 \times 3$ was refined two times resulting in a medium $2000 \times 300 \times 3$ and a fine $4000 \times 600 \times 3$ one.
The \cref{tab:Case-ID_lookup_table}, defines the Case-ID using lattice size, input parameters.
It is followed by \cref{tab:HotKarman_table}, which presents the outcomes for each combination of the kernel, boundary conditions and Case-ID.

\begin{table}[htbp]
\begin{tabular}{lcrrrrrrr}
\toprule
mesh & \multicolumn{1}{c}{\ac{Pr}} & \multicolumn{1}{c}{Lattice size}  & \multicolumn{1}{c}{D} & \multicolumn{1}{c}{U} & \multicolumn{1}{c}{\ac{Re}} & \multicolumn{1}{c}{$\nu$}   & \multicolumn{1}{c}{$k$}  \\ \midrule
coarse &   10 & 1000$\times$150$\times$3 & 30  & 0.01   & 10 & \num{3E-02} & \num{3E-03} \\
medium &   10 & 2000$\times$300$\times$3 & 60  & 0.005  & 10 & \num{3E-02} & \num{3E-03} \\
fine   &   10 & 4000$\times$600$\times$3 & 120 & 0.0025 & 10 & \num{3E-02} & \num{3E-03} \\ \hline
coarse &  100 & 1000$\times$150$\times$3 & 30  & 0.01   & 10 & \num{3E-02} & \num{3E-04} \\
medium &  100 & 2000$\times$300$\times$3 & 60  & 0.005  & 10 & \num{3E-02} & \num{3E-04} \\
fine   &  100 & 4000$\times$600$\times$3 & 120 & 0.0025 & 10 & \num{3E-02} & \num{3E-04} \\ \hline
coarse & 1000 & 1000$\times$150$\times$3 & 30  & 0.01   & 10 & \num{3E-02} & \num{3E-05} \\
medium & 1000 & 2000$\times$300$\times$3 & 60  & 0.005  & 10 & \num{3E-02} & \num{3E-05} \\
fine   & 1000 & 4000$\times$600$\times$3 & 120 & 0.0025 & 10 & \num{3E-02} & \num{3E-05} \\
\bottomrule
\end{tabular}
\caption{The summary of executed cases. The variables, D, U, $\nu$ and $k$ are expressed in lattice units.}
\label{tab:Case-ID_lookup_table}
\end{table}

\begin{table}[htbp]
\newcommand{\fst}{\ensuremath{1^\text{st}}}
\newcommand{\snd}{\ensuremath{2^\text{nd}}}
\newcommand{\bc}{{\footnotesize ord. BC}}
\newcommand{\fstbc}{\multicolumn{1}{c}{\fst}}
\newcommand{\sndbc}{\multicolumn{1}{c}{\snd}}
\begin{tabular}{lcrrrrrrrrrrr}
\toprule
\multirow{2}{*}{mesh} & \multirow[c]{2}{*}{\ac{Pr}} & & \multicolumn{2}{c}{CM-$1^{st}$} & \multicolumn{2}{c}{CM-SRT} & \multicolumn{2}{c}{CM-TRT} & \multicolumn{2}{l}{Cumulants-$1^{st}$} & \multirow{2}{*}{FEM} \\
\cmidrule(lr){3-5}\cmidrule(lr){6-7}\cmidrule(lr){8-9}\cmidrule(lr){10-11}
& & \multicolumn{2}{c}{BC order: \fst} & \sndbc & \fstbc & \sndbc & \fstbc & \sndbc & \fstbc & \sndbc & \\
\midrule
coarse & 10   & &   3.44 &   4.91 &   4.94 &   4.81 &   4.91 &  4.81 &  5.04 &  4.83 &  4.82 \\
medium & 10	  & &   5.03 &   4.84 &   4.87 &   4.81 &   4.86 &  4.81 &  4.92 &  4.82 &  4.82 \\
fine   & 10	  & &   4.92 &   4.83 &   4.84 &   4.81 &   4.83 &  4.81 &  4.87 &  4.81 &  4.82 \\ \midrule
coarse & 100  & &  20.68 &  14.75 &  10.64 &  10.20 &  10.66 & 10.27 & 14.50 & 11.52 & 10.10 \\ 
medium & 100  & &  15.87 &  11.84 &  10.33 &  10.11 &  10.32 & 10.13 & 11.82 & 10.36 & 10.10 \\
fine   & 100  & &  12.96 &  10.64 &  10.20 &  10.09 &  10.19 & 10.08 & 10.83 & 10.13 & 10.10 \\ \midrule
coarse & 1000 & & 166.42 & 102.27 & -71.52 & -57.02 &  27.09 & 24.58 & 94.31 & 58.33 & 21.43 \\
medium & 1000 & & 111.76 &  62.52 &  22.75 &  21.78 &  22.73 & 21.84 & 53.97 & 34.19 & 21.43 \\
fine   & 1000 & &  74.00 &  40.47 &  21.87 &  21.38 &  21.84 & 21.37 & 34.04 & 24.56 & 21.43 \\
\bottomrule
\end{tabular}
\caption{The \ac{Nu} number, computed for different collision kernels and boundary conditions (see~\cref{tab:Case-ID_lookup_table}). \fst order BC -- \ac{BB} (hydrodynamics) \& EQ (thermodynamics), 
\snd order BC -- \ac{IBB} (hydrodynamics) \& \ac{IABB} (thermodynamics).
The \ac{FEM} has been used to obtain reference solution.
}
\label{tab:HotKarman_table}
\end{table}

As far as the numerical conductivity is relatively high ($k=\num{3E-03}$), all kernels provided good results even with first order \ac{BC} on each lattice.
Once the conductivity is lowered by order of magnitude, ($k=\num{3E-04}$), discrepancies occur for the \ac{CM}-$1^{st}$ and Cumulants-$1^{st}$ kernel on the coarse lattice. 
The mismatching results origin from relaxing first-order central moments or cumulants with frequencies corresponding to thermal conductivity, while higher-order quantities were relaxed towards equilibrium. 
Although benefits from the second-order boundary conditions can be clearly observed, they are not sufficient to counterweight the aforementioned effect.
For the most numerically challenging case ($k=\num{3E-05}$), the \ac{CM}-\ac{TRT} kernel provided results with highest quality. 
Interestingly, the \ac{CM}-\ac{SRT} kernel performed reasonably well, except the coarsest lattice for which the wiggles in the temperature field reached the inlet (\cref{fig:HotKarman_Tfield_CM_SRT_numerical_artefacts}), causing the heat flux to be spurious.   
Notice that the behaviour of numerical artefacts for kernel pairs $\lbrace$ \ac{CM}-\ac{SRT}, \ac{CM}-\ac{TRT} $\rbrace$ and $\lbrace$ CM-$1^{st}$, Cumulants-$1^{st}$ $\rbrace$ is similar.
Following the imposed \ac{BC}, the physical temperature range shall be contained within the values of $10$ and $11$.
Other values are undoubtedly artefacts (see \cref{fig:HotKarman_Temperature_field_numerical_artefacts}, in which the temperature range has been clipped to highlight the issue). 
In the case of second-order boundary conditions, the shape of the artefacts was preserved, but the magnitude was decreased.

\newcommand{\cylscale}{0.75}
\begin{figure}[htbp]
        \centering
        \begin{subfigure}[b]{\textwidth}
            \centering
            \includegraphics[width=\cylscale \linewidth]{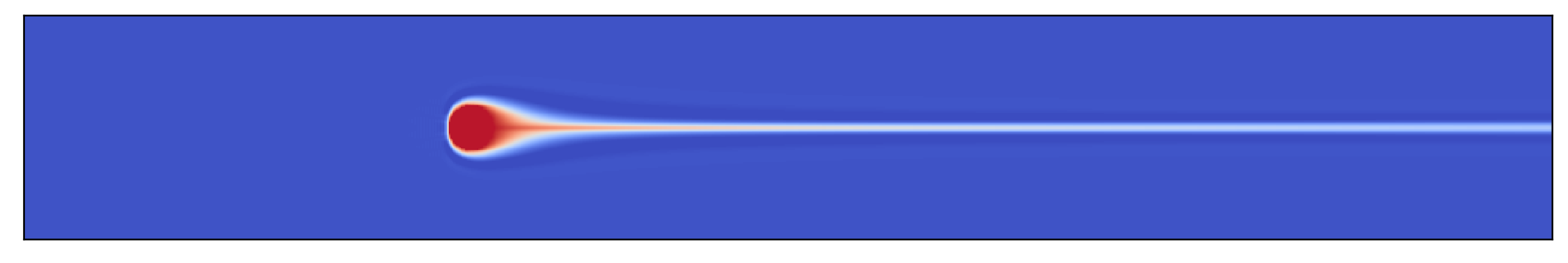}
            \caption{\ac{CM}-$1^{st}$}   
            \label{fig:HotKarman_Tfield_CM}  
       \end{subfigure}
\vskip\baselineskip
        \begin{subfigure}[b]{\textwidth}
            \centering
            \includegraphics[width=\cylscale \linewidth]{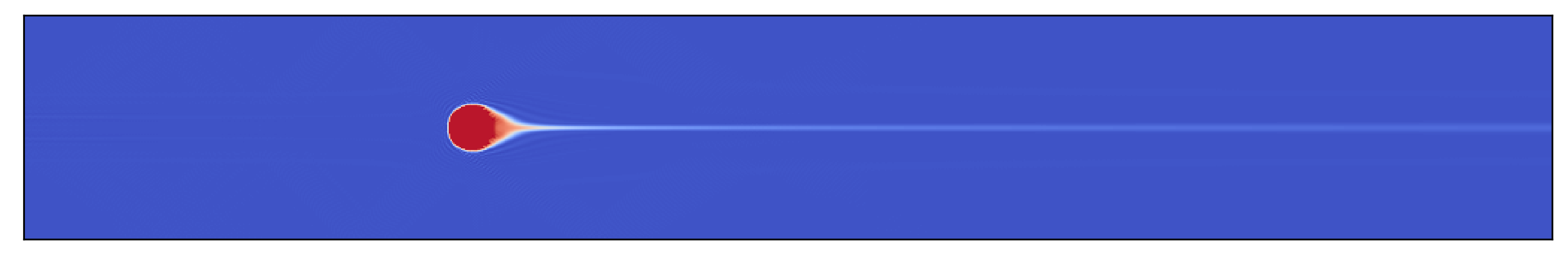}
            \caption{\ac{CM}-\ac{SRT}} 
            \label{fig:HotKarman_Tfield_CM_SRT}    
        \end{subfigure}
\vskip\baselineskip
       \begin{subfigure}[b]{\textwidth}
            \centering
            \includegraphics[width=\cylscale \linewidth]{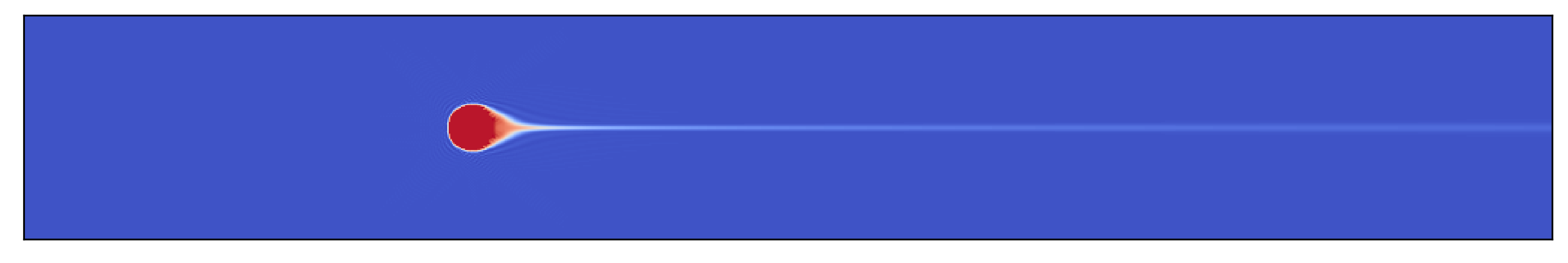}
            \caption{\ac{CM}-\ac{TRT}}   
            \label{fig:HotKarman_Tfield_CM_TRT}  
       \end{subfigure}
\vskip\baselineskip
       \begin{subfigure}[b]{\textwidth}
            \centering
            \includegraphics[width=\cylscale \linewidth]{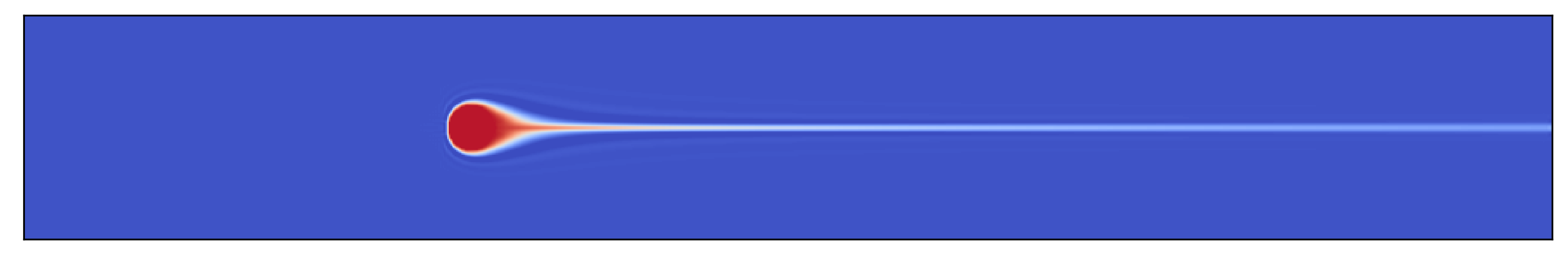}
            \caption{Cumulants-$1^{st}$}  
            \label{fig:HotKarman_Tfield_Cumulants}   
       \end{subfigure}
       %legend
       \begin{subfigure}[b]{\textwidth}
            \centering
            \includegraphics[width=\cylscale \linewidth]{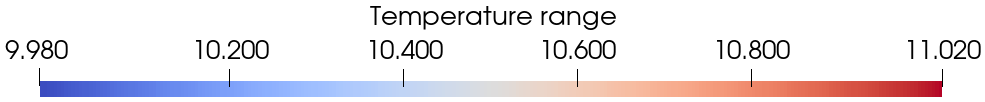}
            \label{fig:HotKarman_T_legend}   
       \end{subfigure}
\centering       
\caption{The temperature field is simulated using four different collision kernels.
The color scale is restricted to $T \in[9.98, 11.02]$.
The flow is computed, on a coarse mesh: 1000$\times$150$\times$3, using $1^\text{st}$ order boundary conditions.
The cylinder diameter is D=30 [lu] and the inlet velocity is U=0.01 [lu/ts]. 
The relaxation frequencies corresponds to $\nu=\num{3e-2}$ and $k=\num{3e-5}$ (\ac{Re} $=10$, \ac{Pr} $=1000$).}
\label{fig:HotKarman_Temperature_field}
\end{figure}

\begin{figure}[htbp]
        \centering
        \begin{subfigure}[b]{\textwidth}
            \centering
            \includegraphics[width=\cylscale \linewidth]{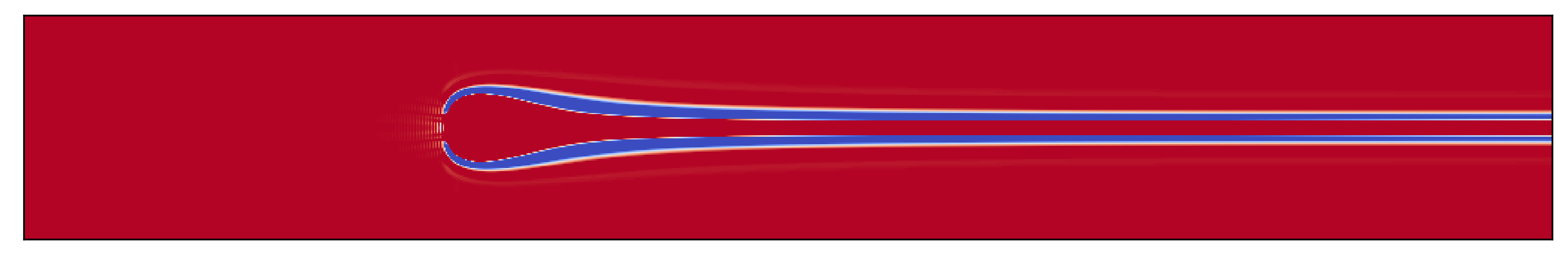}
            \caption{\ac{CM}-$1^{st}$}   
            \label{fig:HotKarman_Tfield_CM_numerical_artefacts}  
       \end{subfigure}
\vskip\baselineskip
        \begin{subfigure}[b]{\textwidth}
            \centering
            \includegraphics[width=\cylscale \linewidth]{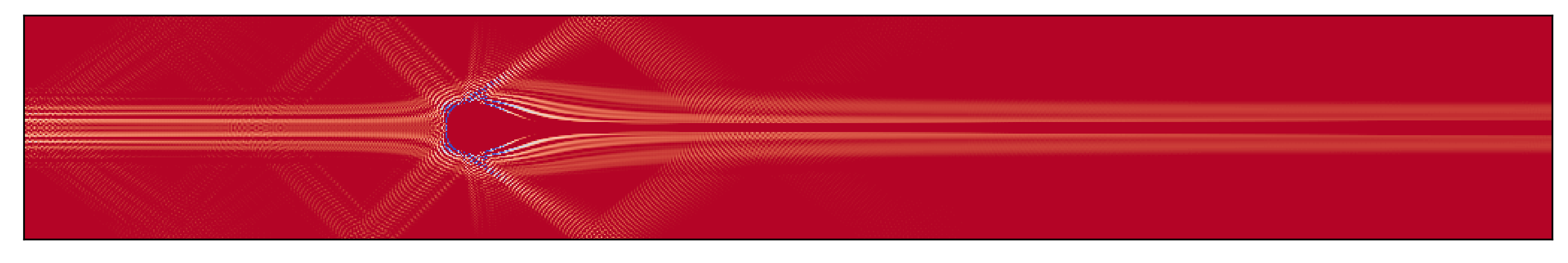}
            \caption{\ac{CM}-\ac{SRT}} 
            \label{fig:HotKarman_Tfield_CM_SRT_numerical_artefacts}    
        \end{subfigure}
\vskip\baselineskip
       \begin{subfigure}[b]{\textwidth}
            \centering
            \includegraphics[width=\cylscale \linewidth]{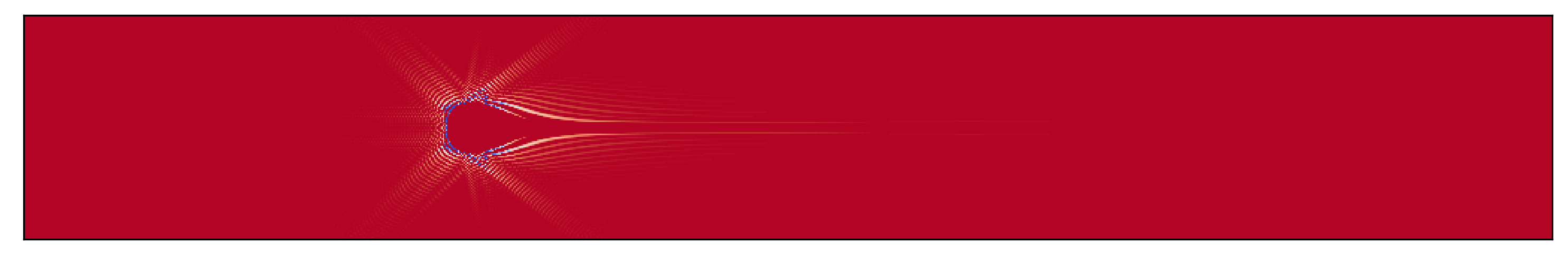}
            \caption{\ac{CM}-\ac{TRT}}   
            \label{fig:HotKarman_Tfield_CM_TRT_numerical_artefacts}  
       \end{subfigure}
\vskip\baselineskip
       \begin{subfigure}[b]{\textwidth}
            \centering
            \includegraphics[width=\cylscale \linewidth]{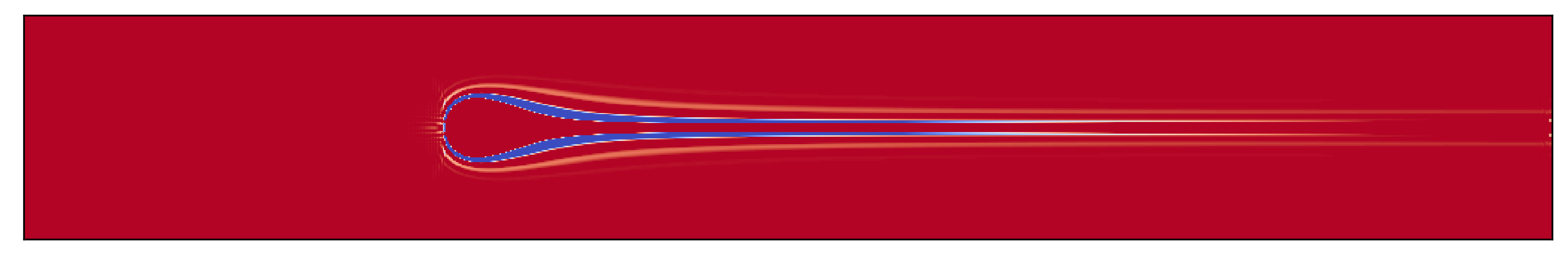}
            \caption{Cumulants-$1^{st}$}  
            \label{fig:HotKarman_Tfield_Cumulants_numerical_artefacts}   
       \end{subfigure}
       %legend
       \begin{subfigure}[b]{\textwidth}
            \centering
            \includegraphics[width=\cylscale \linewidth]{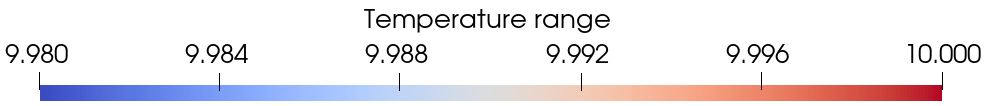}
            \label{fig:HotKarman_T_legend_numerical_artefacts}   
       \end{subfigure}
\centering       
\caption{The temperature field is simulated using four different collision kernels, for the same case as~\ref{fig:HotKarman_Temperature_field}.
Plotted in a color scale restricted the spurious temperature range, $T \in[9.98, 10.00]$, to highlight the numerical artefacts.}
\label{fig:HotKarman_Temperature_field_numerical_artefacts}
\end{figure}

\section{Future outlook}\label{sec:future_outlook}

The authors decided to skip the benchmarks involving buoyancy force (natural convection inside a heated cavity or Rayleigh-Bernard convection) to limit the number of factors which can influence the results. 
Readers interested in the proper treatment of the forcing term are refereed to discussions in~\cite{Guo2002,Geier2015a,fei2017consistent,fei2018threedimensional,Huang2018,DeRosis2019,gruszczynski2020cascaded}.

Although an increase of the resolution might be enough to break the necessity of a larger stencil for the investigated cases, 
it is not obvious whether such an approach would be computationally cheaper, especially that both lattices responsible for the hydrodynamics and the advected field must be refined.

The seek of optimal relaxation frequencies for central moments, or cumulant collision kernel responsible for the advected field deserves further study.
The relationship between higher order cumulants and central moments is non-linear, 
thus the conclusions related to the relaxation of higher-order central moments can not be projected to cumulants in a straightforward fashion.
Due to the similarity between central moments and cumulants for order lower than fourth~\cite{Geier2015a,Coreixas2019}, 
one may expect improvements by relaxing third-order cumulants. 
However, such extension is beyond the scope of the current research.
As a good starting point, research regarding parametrisation of the cumulant kernel responsible for the hydrodynamic field~\cite{Geier2017a} can be recommended.
In terms collision operators which can be expressed in a matrix form, a comprehensive method for recovering equivalent partial differential equations has been recently proposed by~\citet{Fucik2021}.

\section{Conclusions}\label{sec:conclusions}
In this work, an analysis of the state of the art collision kernels applied to both hydrodynamic and the advected field with a focus on the relaxation of the higher-order moments has been presented.
To isolate features which may affects the accuracy of the proposed kernels a set of simple benchmarks was conducted.
The tests has been performed in a numerically challenging, high \acl{Pr} regime.
To alleviate the numerical artefacts raised by low conductivity, a D3Q27 lattice has been utilized for the advected field.
However, lattices with large number of discrete velocities introduce additional degrees of freedom to the relaxation scheme.
We demonstrated that proper treatment of the collision kernel plays a more important role than the application of second-order boundary conditions to represent a curved geometry.
To sum up, for the set of investigated benchmarks, characterized by specific grid resolution and a range of non-dimensional numbers, 
the beneficial effect of tuning of the relaxation coefficients corresponding to the higher order moments has been confirmed numerically
and the \ac{CM}-\ac{TRT} followed by \ac{CM}-\ac{SRT} kernel has been shown to be superior to the kernels that relax only first-order central moments/cumulants.

\section{Acknowledgements}
W. Regulski, J. Szumbarski and M. Dzikowski are appreciated for insightful discussions. C. Leonardi is acknowledged for proofreading of the manuscript.
The computational resources have been provided by 
\textit{Rysy} cluster at Interdisciplinary Centre for Mathematical and Computational Modelling  
and \textit{Prometheus} cluster at Cyfronet Computing Center belonging to PL-Grid Infrastructure.
This work was partially supported by funds granted by QuickerSim Ltd. and Warsaw University of Technology, namely \textit{Rada Naukowa Dyscypliny In\.zynieria Mechaniczna}.
The simulations were completed using the open-source TCLB solver~\cite{Laniewski-Wollk2016a,TCLB} available at: \url{https://github.com/CFD-GO/TCLB}

\section{Conflict of interest}
None declared. 


% \vspace{2em}
%\textbf{Appendices}
\appendix
%\renewcommand{\thesection}{\Alph{section}}
%\setcounter{section}{0}

%\begin{appendices}

\section{Discretisation of the equilibrium distribution function}\label{app:edf_discretization}
Some authors favour the usage of the Hermite polynomials to discretise the equilibrium distribution function~\citep{DeRosisLuo2019,DeRosis2019}.
In this section, an common alternative~\cite{Geier2006} is discussed. 

The continuous Maxwell-Boltzmann distribution function is known as,
\begin{eqnarray}  
\Psi^{\textit{M-B, eq}} = 
\Psi^{\textit{M-B, eq}}(\psi, \boldsymbol{\xi}, \boldsymbol{u}) =
\dfrac{\psi}{(2 \pi c_s^2)^{N/2}} 
exp \left[
-\frac{(\boldsymbol{\xi}-\boldsymbol{u})^2}{2 c_s^2}
\right], \label{eq:Maxwellian}
\end{eqnarray}
where $\psi$ is the quantity of interest (like density or internal energy) and $N$ is the number of spatial dimensions.
The definition of the central moments for a continuous distribution is,
\begin{eqnarray}  
\tilde{\Upsilon}_{mn} = \int_{-\infty}^{\infty} \int_{-\infty}^{\infty} 
(\xi_x - u_x)^m (\xi_y -u_y)^n
\Psi(\psi, \boldsymbol{\xi}, \boldsymbol{u}) 
d \xi_x d \xi_y. \label{eq:cont_cm_mom_def}
\end{eqnarray}
To illustrate the analysis with a relatively short example, a \DQ{2}{9} lattice with the following order of the central moments is investigated,
\begin{linenomath}\begin{align} 
\boldsymbol{\tilde{\Upsilon}} &=
[\tilde{\Upsilon}_{00}, \tilde{\Upsilon}_{10}, \tilde{\Upsilon}_{01}, \tilde{\Upsilon}_{20}, \tilde{\Upsilon}_{02}, \tilde{\Upsilon}_{11}, \tilde{\Upsilon}_{21}, \tilde{\Upsilon}_{12}, \tilde{\Upsilon}_{22}]^\top.
\end{align}\end{linenomath} 
The resulting moments and central moments for the internal energy (with $\psi = H = \Upsilon_{00}^H = \rho c_v T$) are,
\begin{linenomath}\begin{align}
\setstretch{1.4}
\boldsymbol{\Upsilon}^{H, eq} &= 
\Upsilon_{00}^H
\left[\begin{array}{ll}
	1 \\
	u_x \\
	u_y \\
	\left(u_x^{2} + c_s^2\right) \\
	\left(u_y^{2} + c_s^2\right) \\
	u_x u_y \\
	u_y \left(u_x^{2} + c_s^2\right) \\
	u_x \left(u_y^{2} + c_s^2\right) \\
	\left(u_x^{2} u_y^{2} + c_s^2 u_x^{2} + c_s^2 u_y^{2} + c_s^4\right) \\
\end{array}\right],\\
\boldsymbol{\tilde{\Upsilon}}^{H, eq} &= 
\Upsilon_{00}^H
\left[\begin{array}{ll}
	1 \\
	0 \\
	0 \\
	c_s^2 \\
	c_s^2 \\
	0 \\
	0 \\
	0 \\
	c_s^4 \\
\end{array}\right].
\end{align}\end{linenomath}
In general, the central moments of a physical quantity being in equilibrium should be independent of the velocity. 
Next, the discrete equilibrium distribution can be obtained using backward transformation from central moments space,
\begin{linenomath}\begin{align}
\boldsymbol{h}^{eq} = \mathbb{M}^{-1} \mathbb{N}^{-1} \boldsymbol{\tilde{\Upsilon}}^{H, eq}, \label{eq:dedf_full_u}
\end{align}\end{linenomath}
and the result is the same as in case of full order Hermite expansion~\citep{DeRosisLuo2019,DeRosis2019}.
\section{Two Relaxation Time}\label{app:TRT}
To show the origin of the even and odd moments, consider decomposition of the discrete distribution function into symmetric (even) and anti-symmetric (odd) part:
\begin{align*}
h_\alpha^{s}=\frac{h_\alpha + h_{\overline{\alpha}}}{2}, \\
h_\alpha^{a}=\frac{h_\alpha - h_{\overline{\alpha}}}{2}. 
\end{align*}
The discrete equilibrium distribution is treated analogously,
\begin{align*}
h_\alpha^{seq}=\frac{h_\alpha^{eq}+h_{\overline{\alpha}}^{eq}}{2}, \\
h_\alpha^{aeq}=\frac{h_\alpha^{eq}-h_{\overline{\alpha}}^{eq}}{2}. 
\end{align*}
Next, the collision reads,
\begin{linenomath}\begin{align}
\boldsymbol{h}^\star(\textbf{x}, t)&=  \boldsymbol{h} + \frac{1}{\tau_s}[\boldsymbol{h}^{seq} - \boldsymbol{h}^s] + \frac{1}{\tau_a}[\boldsymbol{h}^{aeq} - \boldsymbol{h}^a].
\end{align}\end{linenomath}
Multiplying by $\mathbb{M}$, the scheme can be transferred to the moment space
\begin{linenomath}\begin{align}
\boldsymbol{\Upsilon}^{H,\star}(\textbf{x}, t) &=  \boldsymbol{\Upsilon}^H
+ \frac{1}{\tau_s}[\boldsymbol{\Upsilon}^{H,seq} - \boldsymbol{\Upsilon}^{H,s}] 
+ \frac{1}{\tau_a}[\boldsymbol{\Upsilon}^{H,aeq} - \boldsymbol{\Upsilon}^{H,a}].  \label{eq:trt_in_raw_mom_space}
\end{align}\end{linenomath}
The two-relaxation time approach requires the odd-moments, 
to be relaxed with a common rate $s_{odd}$,
while the even moments, 
with $s_{even}$.  
When the combination of odd and even relaxation rates is kept constant, then the steady state, non-dimensional solution of \ac{NS} or \ac{ADE} is \textit{exactly }controlled by the similarity numbers (Reynolds number, P\'eclet number, etc.)~\cite{Kuzmin2011b}. 
The so called magic parameter has been defined as,
\begin{linenomath}\begin{align}
\Lambda = \left( \dfrac{1}{s_{odd}} - \frac{1}{2} \right) \left( \dfrac{1}{s_{even}} - \frac{1}{2} \right).
\end{align}\end{linenomath}
It is interesting to observe, that vectors describing 
$\left\lbrace \boldsymbol{\Upsilon}^{H,seq}, \boldsymbol{\Upsilon}^{H,s} \right\rbrace $ 
and 
$ \left\lbrace \boldsymbol{\Upsilon}^{H,a}, \boldsymbol{\Upsilon}^{H,a} \right\rbrace$ contain elements with zero values at the same indices (see \cref{eq:raw_mom_seq_aeq,eq:raw_mom_s_a}).
Moreover, the relaxation matrix $ \mathbb{S}^{H}$ is diagonal. 
As a consequence, the \cref{eq:trt_in_raw_mom_space} can be simplified further to:
\begin{linenomath}\begin{align}
\boldsymbol{\Upsilon}^{H,\star}(\textbf{x}, t) 
&= \boldsymbol{\Upsilon}^H + \mathbb{S}^{H} (\boldsymbol{\Upsilon}^{H, eq} - \boldsymbol{\Upsilon}^H)  \nonumber \\ 
&= (\mathbbm{1} - \mathbb{S}^{H}) \boldsymbol{\Upsilon}^H + \mathbb{S}^{H} \boldsymbol{\Upsilon}^{H, eq} .  \label{eq:trt_in_raw_mom_space_simplified}
\end{align}\end{linenomath}
The transformation to the central moments space can be performed in two ways.
In the first approach, which follows the original idea of Ginzburg~\cite{Ginzburg2005,Ginzburg2005a}, the \cref{eq:trt_in_raw_mom_space} is multiplied by $\mathbb{N}$,
\begin{linenomath}\begin{align}
\boldsymbol{\tilde{\Upsilon}}^{H,\star}(\textbf{x}, t) &=  \boldsymbol{\tilde{\Upsilon}}^H
+ \frac{1}{\tau_s}[\boldsymbol{\tilde{\Upsilon}}^{H,seq} - \boldsymbol{\tilde{\Upsilon}}^{H,s}] 
+ \frac{1}{\tau_a}[\boldsymbol{\tilde{\Upsilon}}^{H,aeq} - \boldsymbol{\tilde{\Upsilon}}^{H,a}]. \label{eq:trt_in_raw_cmom_space_Ginzburg}
\end{align}\end{linenomath}
In the second approach, the \cref{eq:trt_in_raw_mom_space_simplified} is transformed,
\begin{linenomath}\begin{align}
\boldsymbol{\tilde{\Upsilon}}^{H,\star}(\textbf{x}, t) 
&= \boldsymbol{\tilde{\Upsilon}}^H + \mathbb{S}^{H} (\boldsymbol{\tilde{\Upsilon}}^{H, eq} - \boldsymbol{\tilde{\Upsilon}}^H)  \nonumber \\ 
&= (\mathbbm{1} - \mathbb{S}^{H}) \boldsymbol{\tilde{\Upsilon}}^H + \mathbb{S}^{H} \boldsymbol{\tilde{\Upsilon}}^{H, eq}. \label{eq:trt_in_raw_cmom_space_simplified}
\end{align}\end{linenomath}
As the structure of $\left\lbrace \boldsymbol{\tilde{\Upsilon}}^{H,seq}, \boldsymbol{\tilde{\Upsilon}}^{H,s} \right\rbrace $ vectors (see \cref{eq:cmom_seq_aeq}) do not contain zero elements,
thus the \cref{eq:trt_in_raw_cmom_space_Ginzburg} does not collapse to \cref{eq:trt_in_raw_cmom_space_simplified}. 

In this contribution, the full order, discrete equilibrium distribution function given in \cref{eq:dedf_full_u} is used to express $h_i^{seq}$ and $h_i^{aeq}$.  Their moments can be calculated using \cref{eq:raw_mom_def,eq:cm_mom_def},
\begin{linenomath}\begin{align}
\setstretch{1.2}
\boldsymbol{\Upsilon}^{H, seq} = 
\Upsilon_{00}^H
\left[\begin{array}{ll}
	1\\
	0 \\
	0 \\
	u_x^{2} + c_s^2 \\
	u_y^{2} + c_s^2 \\
	u_x u_y \\
	0 \\
	0 \\
	u_x^{2} u_y^{2} + c_s^2 u_x^{2} + c_s^2 u_y^{2} + c_s^4\\
\end{array}\right], 
\hspace{3em} 
\boldsymbol{\Upsilon}^{H, aeq} =
\Upsilon_{00}^H
\left[\begin{array}{ll}
	0 \\
	u_x \\
	u_y \\
	0 \\
	0 \\
	0 \\
	u_y \left(u_x^{2} + c_s^2\right) \\
	u_x \left(u_y^{2} + c_s^2\right) \\
	0 
\end{array}\right]. \label{eq:raw_mom_seq_aeq}
\end{align}\end{linenomath}
The moments of symmetric and anti-symmetric distribution function exhibits analogous structure,
\begin{linenomath}\begin{align}
\setstretch{1.2}
\boldsymbol{\Upsilon}^{H, s} = 
\left[\begin{array}{ll}
	\Upsilon_{00}^{H,s} \\
	0 \\
	0 \\
	\Upsilon_{20}^{H,s} \\
	\Upsilon_{02}^{H,s} \\
	\Upsilon_{11}^{H,s} \\
	0 \\
	0 \\
	\Upsilon_{22}^{H,s} \\
\end{array}\right],
\hspace{3em}
\boldsymbol{\Upsilon}^{H, a}= 
\left[\begin{array}{ll}
	0 \\
	\Upsilon_{10}^{H,s} \\
	\Upsilon_{01}^{H,s} \\
	0 \\
	0 \\
	0 \\
	\Upsilon_{21}^{H,s} \\
	\Upsilon_{12}^{H,s} \\
	0 
\end{array}\right]. \label{eq:raw_mom_s_a}
\end{align}\end{linenomath}
These results extend the analysis done in~\cite{Lu2019}, which was limited to raw, second-order moments and truncated (to linear and quadratic velocity terms) discrete equilibrium function. 
Finally, the central moments are,
\begin{linenomath}\begin{align}
\setstretch{1.2}
\boldsymbol{\tilde{\Upsilon}}^{H, seq} = 
\Upsilon_{00}^H
\left[\begin{array}{ll}
	1 \\
	- u_x \\
	- u_y \\
	2 u_x^{2} + c_s^2 \\
	2 u_y^{2} + c_s^2 \\
	2 u_x u_y \\
	- u_y \left(4 u_x^{2} + c_s^2\right) \\
	- u_x \left(4 u_y^{2} + c_s^2\right) \\
	8 u_x^{2} u_y^{2} + 2 c_s^2 (u_x^{2} + u_y^{2}) + c_s^4 \\
\end{array}\right],
\hspace{3em}
\boldsymbol{\tilde{\Upsilon}}^{H, aeq} = 
\Upsilon_{00}^H
\left[\begin{array}{ll}
	0 \\
	u_x \\
	u_y \\
	- 2 u_x^{2} \\
	- 2 u_y^{2} \\
	- 2 u_x u_y \\
	u_y \left(4 u_x^{2} + c_s^2\right) \\
	u_x \left(4 u_y^{2} + c_s^2\right) \\
	- \left(8 u_x^{2} u_y^{2} + 2 c_s^2 (u_x^{2} + u_y^{2}) \right) \\
\end{array}\right]. \label{eq:cmom_seq_aeq}
\end{align}\end{linenomath}

\section{Transformation Matrices}\label{app:NM_matrices}
From the computational point of view, it is preferred to perform the transformations to the central moments space in two steps, using  \cref{eq:to_raw_mom,eq:to_cmom}.
Since $ \boldsymbol{\tilde{\Upsilon}}  = \mathbb{T} \boldsymbol{g} = \mathbb{N} \mathbb{M} \boldsymbol{g}$, the $\mathbb{N}$ matrix can be found as $\mathbb{N} = \mathbb{T} \mathbb{M}^{-1} $.
Rows of the transformation matrices are calculated analogously to $\Upsilon$ and $\tilde{\Upsilon}$.  
For a \DQ{3}{27} lattice, each row consists of $q \in \left\lbrace 1,2,...,27 \right\rbrace$ elements, 
\begin{linenomath}\begin{align}
 \boldsymbol{M}_{mno} &= \left[ \left.M_{mno}\right|_1,  \left.M_{mno}\right|_2,  ...,  \left.M_{mno}\right|_\alpha, ..., \left.M_{mno}\right|_{q-1}, \left.M_{mno}\right|_q \right], \\
 \boldsymbol{T}_{mno} &= \left[ \left.T_{mno}\right|_1,  \hspace{0.35em} \left.T_{mno}\right|_2,  \hspace{0.35em} ...,  \left.T_{mno}\right|_\alpha, \hspace{0.35em} ..., \left.T_{mno}\right|_{q-1}, \hspace{0.35em} \left.T_{mno}\right|_q \right]. 
\end{align}\end{linenomath}
The $mno$ subscript refers to the order of moment, while $\alpha -th$ subscript indicates the index of the element in the $\boldsymbol{M}_{mno}$ row.
Each element can be calculated as,
\begin{linenomath}\begin{align}
\left.M_{mno}\right|_\alpha &= (e_{\alpha x})^m (e_{\alpha y})^n ( e_{\alpha z})^o,  \label{eq:M} \\
\left.T_{mno}\right|_\alpha &= ( e_{\alpha x} - u_x)^m ( e_{\alpha y} - u_y)^n ( e_{\alpha z} - u_z)^o. \label{eq:T}
\end{align}\end{linenomath}
Next, the matrices are assembled row by row as,
\begin{linenomath}\begin{align*}
\mathbb{M}  
 &= 
 \left[
 \boldsymbol{M}_{000}, 
 \boldsymbol{M}_{100}, 
 \boldsymbol{M}_{010}, 
 \boldsymbol{M}_{001}, 
 \boldsymbol{M}_{110},
 \boldsymbol{M}_{101},
 \boldsymbol{M}_{011},
 \boldsymbol{M}_{200},
 \boldsymbol{M}_{020},
 \boldsymbol{M}_{002},
 ...\boldsymbol{M}_{ijk}...,
 \boldsymbol{M}_{122},
 \boldsymbol{M}_{212},
 \boldsymbol{M}_{221},
 \boldsymbol{M}_{222}
 \right]^\top,
  \\
\mathbb{T} &= 
 \left[
 \boldsymbol{T}_{000}, \hspace{0.35em}
 \boldsymbol{T}_{100}, \hspace{0.35em}
 \boldsymbol{T}_{010}, \hspace{0.35em}
 \boldsymbol{T}_{001}, \hspace{0.35em}
 \boldsymbol{T}_{110}, \hspace{0.35em}
 \boldsymbol{T}_{101}, \hspace{0.35em}
 \boldsymbol{T}_{011}, \hspace{0.35em}
 \boldsymbol{T}_{200}, \hspace{0.35em}
 \boldsymbol{T}_{020}, \hspace{0.35em}
 \boldsymbol{T}_{002}, \hspace{0.35em}
 ...\boldsymbol{T}_{ijk}..., \hspace{0.35em}
 \boldsymbol{T}_{122}, \hspace{0.35em}
 \boldsymbol{T}_{212}, \hspace{0.35em}
 \boldsymbol{T}_{221}, \hspace{0.35em}
 \boldsymbol{T}_{222} \hspace{0.3em}
 \right]^\top.
\end{align*}\end{linenomath}

\section{Source term treatment}\label{app:source_term}
For completeness of the study, the addition of the source term to the cumulant collision kernel is described in this appendix from the theoretical point of view. 
It is known in the literature~\cite{Kruger2017}, that the integration of the discrete Boltzmann equation with trapezoidal rule leads to implicit evolution equation,
\begin{linenomath}\begin{align}
h_\alpha(\mathbf{x} + \mathbf{e}_\alpha\delta t, t + \delta t ) &= 
\Omega_{\text{H},\alpha} \left(\mathbf{h}(\mathbf{x}, t) \right)
+ \frac{1}{2}\Big[q_\alpha(\mathbf{x} + \mathbf{e}_\alpha\delta t, t + \delta t ) + q_\alpha (\mathbf{x}, t) \Big].
\end{align}\end{linenomath}
To remove the implicitness, a shift of variables is conducted and denoted with tilde,
\begin{linenomath}\begin{align}
\tilde h_\alpha  &= h_\alpha - \dfrac{1}{2} q_\alpha \\
\tilde H &= \sum_\alpha \tilde h_\alpha = H - \dfrac{1}{2}Q.
\end{align}\end{linenomath}
The $q_\alpha$ and $Q$ denote the discretized source term and its zeroth moment respectively.
A more detailed discussion of this procedure can be found in the recent work done by authors~\cite{GGMDLLW_source_term_arxiv2021} and references therein.

By transforming source term to the cumulant space, where the collision is performed, the augmented form of~\cref{eq:H_cu_collision} can be written as,
\begin{linenomath}\begin{align}
% cu internal energy field
\Omega_\text{H}(\mathbf{\tilde{h}}) &= \mathscr{C}^{-1}\bigg(\boldsymbol{\tilde{\mathcal{C}}}^{\text{H}} 
+ \mathbb{S}^{\text {H}} \Big(\boldsymbol{\tilde{\mathcal{C}}}^{\text{H,eq}} - \boldsymbol{\tilde{\mathcal{C}}}^{\text{H}} \Big) 
+ \Big( 1 - \mathbb{S}^{\text {H}}/2 \Big) \boldsymbol{\mathcal{C}}^{Q} \bigg).
\label{eq:H_cu_collision_with_Q}
\end{align}\end{linenomath}
The simplicity of the source term in the cumulants space is noteworthy,
\begin{linenomath}\begin{align}
\boldsymbol{\mathcal{C}}^{Q} 
=
\mathscr{C}(Q)
= 
& \left[
 c_{000}^{Q}, 
 c_{100}^{Q}, 
 c_{010}^{Q}, 
 c_{001}^{Q}, 
 c_{110}^{Q},
 c_{101}^{Q},
 c_{011}^{Q},
 c_{200}^{Q},
 c_{020}^{Q},
 c_{002}^{Q},
 ...c_{ijk}^{eq}...,
 c_{122}^{Q},
 c_{212}^{Q},
 c_{221}^{Q},
 c_{222}^{Q}
 \right]^\top  \nonumber \\
 = 
& \left[
 Q, \hspace{0.8em}
 0, \hspace{1.05em}
 0, \hspace{1.05em}
 0, \hspace{1.05em}
 0, \hspace{1.05em}
 0, \hspace{1.05em}
 0, \hspace{1.05em} 
 0, \hspace{1.05em}
 0, \hspace{1.05em}
 0, \hspace{1.05em}
  ...0..., \hspace{0.8em} 
 0, \hspace{1.05em} 
 0, \hspace{1.05em}
 0, \hspace{1.05em} 
 0 \hspace{1.0em} 
 \right]^\top.
\end{align}\end{linenomath}

%\newpage
%\textbf{References}
%
\bibliographystyle{model1-num-names}
\bibliography{elsarticle-template-CuLBM.bib}






% \end{thebibliography}

\end{document}